\documentclass[review,12pt]{elsarticle}

\usepackage{amssymb}
\usepackage{amsthm}
\usepackage{amsmath}
\usepackage{mathrsfs}
\usepackage{graphicx}
\usepackage{epstopdf}
\usepackage{float}
\usepackage{caption}
\usepackage{subcaption}
\usepackage{bm}
\usepackage{physics}
\usepackage{bm}
\usepackage{bbm}
\usepackage{mathrsfs}
\usepackage{cleveref}
\usepackage{soul}
\usepackage{booktabs}
\usepackage{longtable}
\usepackage{amssymb}
\usepackage{xcolor}
\newcommand{\greencheck}{{\color{green}\checkmark}}
\newcommand{\redmark}{{\color{red}$\boldsymbol{\times}$}}
\usepackage[margin=2cm]{geometry}
\biboptions{sort&compress}

\journal{International Journal of Hydrogen Energy}

\makeatletter
\def\@author#1{\g@addto@macro\elsauthors{\normalsize%
    \def\baselinestretch{1}%
    \upshape\authorsep#1\unskip\textsuperscript{%
      \ifx\@fnmark\@empty\else\unskip\sep\@fnmark\let\sep=,\fi
      \ifx\@corref\@empty\else\unskip\sep\@corref\let\sep=,\fi
      }%
    \def\authorsep{\unskip,\space}%
    \global\let\@fnmark\@empty
    \global\let\@corref\@empty  
    \global\let\sep\@empty}%
    \@eadauthor={#1}
}
\makeatother

\begin{document}

\begin{frontmatter}



\title{Computational modelling of hydrogen assisted fracture in polycrystalline materials}


\author[IMT,US]{A. Valverde-Gonz\'alez}

\author[IC]{E. Mart\'{\i}nez-Pa\~neda\corref{cor1}}
\ead{e.martinez-paneda@imperial.ac.uk}

\author[IC,UDG]{A. Quintanas-Corominas}

\author[US]{J. Reinoso}

\author[IMT]{M. Paggi}

\address[IMT]{IMT School for Advanced Studies, Piazza San Francesco 19, Lucca 55100, Italy}

\address[US]{Grupo de Elasticidad y Resistencia de Materiales, Escuela Técnica Superior de Ingeniería, Universidad de Sevilla, Camino de los Descubrimientos s/n, 41092 Sevilla, Spain}

\address[IC]{Department of Civil and Environmental Engineering, Imperial College London, London SW7 2AZ, UK}

\address[UDG]{AMADE, University of Girona, Polytechnic School, c/Universitat de Girona 4, 17003 Girona, Spain}

\cortext[cor1]{Corresponding author.}

\begin{abstract}
We present a combined phase field and cohesive zone formulation for hydrogen embrittlement that resolves the polycrystalline microstructure of metals. Unlike previous studies, our deformation-diffusion-fracture modelling framework accounts for hydrogen-microstructure interactions and explicitly captures the interplay between bulk (transgranular) fracture and intergranular fracture, with the latter being facilitated by hydrogen through mechanisms such as grain boundary decohesion. We demonstrate the potential of the theoretical and computational formulation presented by simulating inter- and trans-granular cracking in relevant case studies. Firstly, verification calculations are conducted to show how the framework predicts the expected qualitative trends. Secondly, the model is used to simulate recent experiments on pure Ni and a Ni-Cu superalloy that have attracted particular interest. We show that the model is able to provide a good quantitative agreement with testing data and yields a mechanistic rationale for the experimental observations.\\
\end{abstract}


\end{frontmatter}



\section{Introduction}
\label{Introduction}

Hydrogen is at the core of the most promising solutions to our energy crisis. Hydrogen isotopes fuel the nuclear fusion reaction, the most efficient potentially useable energy process. Moreover, hydrogen is widely seen as the energy carrier of the future and the most versatile means of energy storage. It can be produced via electrolysis from renewable sources, such as wind or solar power, and stored to be used as a fuel or as a raw material in the chemical industry. Hampering these opportunities, hydrogen is known for causing catastrophic failures in metallic structures. Through a phenomenon often termed hydrogen embrittlement, metals exposed to hydrogen containing environments experience a significant reduction in ductility, fracture toughness and fatigue resistance \cite{Gangloff2003,Gangloff2012,Djukic2019}. In the presence of hydrogen, otherwise ductile metals fail in a brittle manner, with cracking often nucleating and propagating along grain boundaries \citep{Harris2018,Nagao2018}. This ductile-to-brittle shift of metallic alloys in hydrogeneous environments is arguably one of the biggest threats to the deployment of a hydrogen energy infrastructure.\\

There is a vast literature devoted to shed light into the physical mechanisms behind hydrogen embrittlement \cite{Dadfarnia2010,AM2016,Katzarov2017a,Lynch2019,Yu2020a,Shishvan2020}, and to develop mechanistic predictive models that can prevent failures and map safe regimes of operation \cite{Olden2012,Matsumoto2017,Diaz2017,TAFM2020c,IJP2021}. The vast majority of the computational models developed for predicting hydrogen assisted cracking fall into two categories: (i) discrete methods, such as cohesive zone models \cite{Serebrinsky2004,Moriconi2014,Yu2016a,EFM2017}, and (ii) diffuse approaches, such as phase field or other non-local damage models \cite{CMAME2018,Duda2018,Anand2019,JMPS2020,Huang2020,Wu2020b}. Cohesive zone models and other discrete methodologies are suitable to describe the nucleation and propagation of sharp cracks through a predefined path. Phase field fracture methods have additional modelling capabilities and can also deliver predictions when the crack trajectory is unknown, when failure is triggered by defects of arbitrary shape, and when the fracture process is complex (e.g., involving the interaction between multiple defects). Both kinds of models can be readily coupled with the hydrogen transport equation and have been successful in qualitatively capturing the main experimental trends, such as the sensitivity to loading rate, hydrogen concentration, and material strength. However, these modelling studies treat materials as isotropic continuum solids, without resolving the underlying microstructure. A number of icrostructurally-sensitive works have been recently carried out (see, e.g.,  \cite{Castelluccio2018,Hassan2019,Kumar2020,Ogosi2020,Hussein2021,Tondro2022,Das2022} and Refs. therein), but these are limited to capture the interplay between diffusion and deformation, and do not explicitly simulate fracture. The micromechanical fracture of polycrystalline materials in hydrogen-containing environments has been simulated in the works of Rimoli and Ortiz \cite{Rimoli2010}, Benedetti \textit{et al.} \cite{Benedetti2018}, and De Francisco \textit{et al.} \cite{DeFrancisco2022}. In these works, a cohesive zone formulation was used to predict the failure of grain boundaries, neglecting transgranular cracks.\\

In this work, we present a new microstructurally-sensitive computational framework for predicting hydrogen assisted fractures. For the first time, the model combines a phase field description of bulk fracture with a cohesive zone model for intergranular cracking. This enables capturing both ductile transgranular fracture and brittle intergranular fracture, and the transition from one to the other. The mechanical and hydrogen transport problems are strongly coupled, with the hydrostastic stress driving hydrogen transport and the hydrogen content reducing the grain boundary strength. The fracture of polycrystalline solids is simulated, with the bulk deformation response being characterised by von Mises plasticity theory. Numerical experiments are conducted to gain insight into the mechanisms of hydrogen-assisted grain boundary decohesion. Focus is on Ni and Ni superalloys, where hydrogen assisted failures are known to be governed by grain boundary decohesion \cite{Harris2018,Tehranchi2019}. Among other case studies, the model is used to provide a mechanistic rationale to two recent sets of experiments on Monel K-500 \cite{CS2020} and pure Ni \cite{Harris2018} that have attracted particular interest in the hydrogen embrittlement community. 

\section{Modelling framework}
\label{Sec:Form}

\subsection{Fundamentals}
\label{SubSec:Introduction}

Our model deals with an arbitrary body $\Omega \in \mathbb{R}^{n_{dim}}$ with a delimiting external surface $\partial \Omega \in \mathbb{R}^{n_{dim}-1}$ of outward normal \textbf{n}. The body $\Omega$ contains a discrete internal discontinuity $\Gamma$ associated with fracture events in the bulk, and also pre-existing interfaces, arranged in the set $\Gamma_i$. Through exposure to a hydrogen containing environment, hydrogen ingress takes place and thus hydrogen atoms can diffuse through the body, driven by gradients of chemical potential. As detailed in the forthcoming subsections, this leads to a three-field boundary value problem, where the displacement field, the fracture status, and the hydrogen concentration are the primal kinematic variables.\\ 

We build our framework upon the assumption of small strains. The vector $\mathbf{x}$ is used to denote the position of an arbitrary point in the global Cartesian system. From a mechanical perspective, the delimiting surface of the body is decomposed into two regions; one where the displacements \textbf{u} are prescribed through Dirichlet-type boundary conditions (BCs), $\partial \Omega_{u}$, and one where tractions $\textbf{t}$ are prescribed \textit{via} Neumann-type BC, such that $\partial \Omega = \partial \Omega_{u} \cup  \partial \Omega_{t} $ and $ \partial \Omega_{u} \cap  \partial \Omega_{t}  = \emptyset $.  The deformation process is characterized by the small deformation tensor $\boldsymbol{\varepsilon} (\mathbf{x})$, which is defined as the symmetric part of the displacement gradient: $\boldsymbol{\varepsilon} (\mathbf{x}) : = \nabla^{s} \mathbf{u} (\mathbf{x}) $. Bulk fracture phenomena is here captured using the phase field fracture method \cite{Bourdin2000,PTRSA2021}. Hence, crack discontinuities are regularised within a diffuse region, whose thickness is characterised by a phase field length scale $\ell$, and the evolution of the crack-solid interface is described by an auxiliary phase field variable $\phi$. Regarding hydrogen transport, the external body surface is divided into two parts: the region $\partial \Omega_{q}$, where the hydrogen flux $\mathbf{J}$ can be prescribed through a Neumann-type BC, and the region $\partial \Omega_{C}$ where the hydrogen concentration $C$ can be prescribed using a Dirichlet BC.\\

\begin{figure}[t]
	\centering
	\includegraphics[width=1\textwidth]{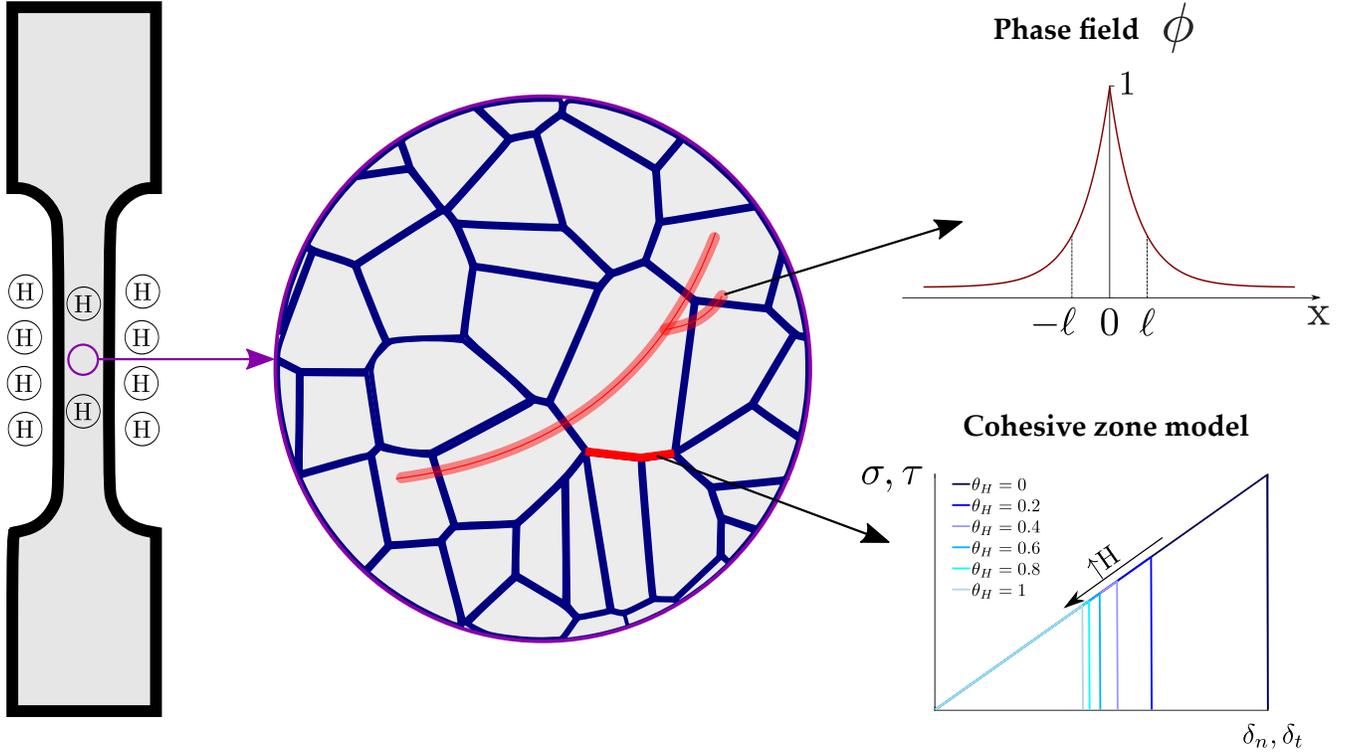}
	\caption{Sketch summarising the modelling framework: the polycrystalline microstructure of metals exposed to hydrogen is explicitly simulated, with the phase field method being employed to describe the growth of ductile transgranular cracks while hydrogen-assisted grain boundary decohesion is captured by means of a cohesive zone model. Some of the key variables of the model are shown; namely, the phase field order parameter $\phi$, the phase field length scale $\ell$ (governing the size of the interface region), the hydrogen coverage $\theta_H$, and the cohesive tractions ($\sigma, \tau$) and separations ($\delta_n, \delta_t$).}
	\label{fig:Sketch} 
\end{figure}

In addition to bulk fracture, our microstructurally-sensitive formulation employs a cohesive zone model to describe the failure of grain boundaries. As shown in Fig. \ref{fig:Sketch}, we explicitly model the polycrystalline microstructure of metals and predict intergranular and transgranular cracking events. The phase field fracture method is used to describe transgranular cracks, which are associated with a ductile fracture process, while a cohesive zone model is employed to predict the decohesion of grain boundaries. Since the latter correspond to a brittle failure and are triggered by the presence of hydrogen, the interfacial fracture energy $\gamma_c$ is defined as a function of the hydrogen concentration $C$. Conversely, the bulk material toughness $G_c$ is considered to be a hydrogen-insensitive constant that characterises the resistance of the material to undergo microvoid cracking. Accordingly, for a solid with strain energy density $\psi(\boldsymbol{\varepsilon},\phi)$, the internal functional of the mechanical system comprises the bulk ($\Pi^{(b)}_{int}$) and interfacial ($\Pi^{(i)}_{int}$) contributions, reading
\begin{equation}
\label{eq:2.1.1}
\Pi_{int}(\mathbf{u},\phi,C) = \underbrace{\int_{\Omega/ \Gamma_l}\psi(\boldsymbol{\varepsilon},\phi)\, \mathrm{d} V + \int_{\Gamma_l}G_{c} \, \mathrm{d}\Gamma}_{\Pi^{(b)}_{int}} + \underbrace{\int_{\Gamma}\gamma_c(C) \, \mathrm{d}\Gamma_i}_{\Pi^{(i)}_{int}}
\end{equation}

In the subsequent subsections we proceed to describe each of the features and the constitutive choices of our micromechanical model.

\subsection{Chemo-elastoplasticity}
\label{SubSec:HydrogenDiff}

The deformation and diffusion problems are intrinsically coupled as hydrogen transport within the crystal lattice is driven by gradients of concentration and hydrostatic stress. We focus our attention on the transport of diffusible hydrogen and consider the influence of traps in the cracking process by using the Langmuir-Mclean isotherm to estimate the hydrogen coverage at grain boundaries. The role of microstructural traps upon slowing down diffusion can also be taken into account through an appropriate choice of the (apparent) diffusion coefficient $D$. Mass conservation requirements relate the rate of change of the hydrogen concentration $C$ to the hydrogen flux through the external surface,
\begin{equation}
    \int_{\Omega} \frac{\text{d}C}{\text{d}t} \, \mathrm{d}V + \int_{\partial \Omega} \mathbf{J} \cdot \mathbf{n} \, \mathrm{d}S = 0
\end{equation}

The strong form of the balance equation can be readily obtained by making use of the divergence theorem and noting that the expression must hold for any arbitrary volume,
\begin{equation}\label{Eq:StrongC}
    \frac{\text{d}C}{\text{d}t} + \nabla \cdot \mathbf{J} = 0
\end{equation}

\noindent For an arbitrary, suitably continuous, scalar field, $\delta C$, the variational statement (\ref{Eq:StrongC}) reads,
\begin{equation}
    \int_\Omega \delta C \left( \frac{\text{d}C}{\text{d}t} + \nabla \cdot \mathbf{J} \right) \, \mathrm{d}V  = 0
\end{equation}

\noindent Rearranging, and making use of the divergence theorem, the weak form renders,
\begin{equation}
   \int_\Omega \left[ \delta C \left( \frac{\text{d}C}{\text{d}t} \right) - \mathbf{J} \cdot \nabla \delta C   \right] \, \mathrm{d}V + \int_{\partial \Omega_q} \delta C q \, \mathrm{d}S  = 0 
\end{equation}

\noindent where $q=\mathbf{J} \cdot \mathbf{n}$ is the concentration flux exiting the body across $\partial \Omega_q$. The diffusion is driven by the gradient of the chemical potential $\nabla \mu$, with the chemical potential of hydrogen in lattice sites being given by,
\begin{equation}\label{Eq:ChePotential}
    \mu=\mu^0 + RT \ln \frac{\theta_L}{1-\theta_L} - \bar{V}_H \sigma_H
\end{equation}

\noindent Here, $\theta_L$ is the occupancy of lattice sites, which is related to the concentration and number of sites as $\theta_L=C/N$. Also, $\mu^0$ denotes the chemical potential in the standard case, and $\sigma_H$ is the hydrostatic stress. The last term of (\ref{Eq:ChePotential}) corresponds to the so-called stress-dependent part of the chemical potential $\mu_{\sigma}$, with $\bar{V}_H$ being the partial molar volume of hydrogen in solid solution. The mass flux follows a linear Onsager relationship with $\mu_{\sigma}$, which is often derived from (\ref{Eq:ChePotential}) by adopting the assumptions of low occupancy ($\theta_L \ll 1$) and constant interstitial sites concentration ($\nabla N=0$) (see, e.g., \cite{Alvaro2014c,AM2020}), such that
\begin{equation}\label{Eq:Flux}
    \mathbf{J}= - \frac{D C}{R T} \nabla \mu = -D \nabla C + \frac{D}{RT} C \bar{V}_H \nabla \sigma_H
\end{equation}

\noindent where $T$ is the absolute temperature and $R$ is the ideal gas constant. Accordingly, the hydrogen transport equation becomes
\begin{equation}\label{Eq:WeakC}
    \int_\Omega \left[ \delta C \left( \frac{1}{D} \frac{dC}{dt} \right) + \nabla \delta C  \nabla C  - \nabla \delta C \left( \frac{\bar{V}_H C}{RT}  \nabla \sigma_H \right) \right] \, \mathrm{d}V = - \frac{1}{D} \int_{\partial \Omega_q} \delta C q \, \mathrm{d}S
\end{equation}

As evident from (\ref{Eq:ChePotential})-(\ref{Eq:WeakC}), the presence of hydrostatic stresses (or volumetric strains) brings a reduction in chemical potential and an increase in hydrogen solubility. Thus, an appropriate description of the stress state in the solid is needed to quantitatively estimate the hydrogen distribution. Here, we choose to describe the deformation of the solid using conventional von Mises plasticity. Accordingly, the total strain energy density of the solid is given by the sum of the elastic and plastic parts,
\begin{equation}\label{eq:StrainEnergyDensity}
    \psi = \psi^e \left(\bm{\varepsilon}^e \right) + \psi^p \left( \bm{\varepsilon}^p \right)= \frac{1}{2} \lambda \left[ \text{tr} \left( \bm{\varepsilon}^e \right) \right]^2 + \mu \, \text{tr} \left[  \left( \bm{\varepsilon}^e \right)^2 \right] + \int_0^t \left( \bm{\sigma} : \dot{\bm{\varepsilon}}^p \right) \text{d} t \, .
\end{equation}

\noindent where $\lambda$ is the first Lame parameter and $\mu$ is the shear modulus. Also, the Cauchy stress tensor is given by $\bm{\sigma}$, and $\bm{\varepsilon}^e$ and $\bm{\varepsilon}^p$ respectively denote the elastic and plastic strain tensors. Isotropic power law hardening behaviour is assumed by adopting the following hardening law:
\begin{equation}
\sigma_f=\sigma_y\left( 1+\frac{E\varepsilon^p}{\sigma_y} \right)^{(1/n)}
\label{eq:2.2.10}
\end{equation}

\noindent where $\sigma_f$ and $\sigma_y$ are the current and initial yield stresses, $E$ is Young's modulus, $\varepsilon^p$ is the accumulated equivalent plastic strain and $n$ is the strain hardening exponent. One should note that, for simplicity, we have chosen to neglect the role of plastic strain gradients; however, large plastic strain gradients arise in the vicinity of cracks and other stress concentrators and lead to large crack tip tensile stresses and hydrogen concentrations \cite{IJHE2016,TAFM2017}. The extension of the present framework to account for the role of plastic strain gradients and geometrically necessary dislocations (GNDs) will be addressed in future work. 

\subsection{A phase field description of transgranular fractures}
\label{SubSec:PhaseField}

The phase field fracture method is used to regularise the internal discontinuity $\Gamma_\ell$, representing the nucleation and growth of transgranular (ductile) cracks. An auxiliary phase field variable $\phi(\mathbf{x})$ is used to describe the evolution of the solid-crack interface, taking the value of $\phi=0$ for the pristine state and of $\phi=1$ for the fully damaged state. Phase field fracture methods have gained remarkable popularity in recent years and have been successfully used to predict cracking in a wide range of materials and applications, including functionally graded solids \cite{CPB2019,Kumar2021}, shape memory alloys \cite{CMAME2021,FFEMS2022}, and fibre-reinforced composites \cite{Carollo2017,CST2021}. The evolution of the phase field equation is dictated by the energy balance associated with the thermodynamics of fracture, as first presented by Griffith \cite{Griffith1920} and later extended to elastic-plastic solids by Orowan \cite{Orowan1948}. Thus, under prescribed displacements, the variation of the total energy $\Pi$ due to an incremental increase of the crack area d$A$ is given by,
\begin{equation}\label{eq:Griffith}
    \frac{\text{d}\Pi}{\text{d}A} = \frac{\text{d} \Psi }{\text{d}A} + \frac{\text{d} W_c}{\text{d}A} = 0
\end{equation}

\noindent where $W_c$ is the work required to create two new surfaces and $\Psi=\int \psi \, \text{d}V$ is the stored strain energy. The last term in Eq. (\ref{eq:Griffith}) denotes the material toughness $G_c=\text{d}W_c/\text{d}A$, which can be as low as some tens of J/m$^2$ for brittle solids or as high as thousands of kJ/m$^2$ for ductile metals where plastic dissipation enhances fracture resistance. The energy balance of Eq. (\ref{eq:Griffith}) can be cast in a variational form as \cite{Francfort1998}:
\begin{equation}\label{eq:Francfort}
  \Pi = \int_\Omega \psi \left( \bm{\varepsilon} \right)  \, \text{d} V + \int_{\Gamma} G_c \, \text{d} \Gamma
\end{equation}

The energy balance is now global and fracture phenomena can be predicted by minimizing the total energy $\Pi$. However, minimization of Eq. (\ref{eq:Francfort}) is hindered by the unknown nature of the discontinuous crack surface $\Gamma$. The phase field regularization can then be exploited to smear this sharp interface into a diffuse region, whose thickness is governed by a phase field length scale $\ell$. Accordingly, the energy balance (\ref{eq:Francfort}) can be approximated as \cite{Bourdin2000}:
\begin{equation}\label{eq:Potential}
    \Pi_\ell = \int_\Omega g \left( \phi \right) \psi_0 \left( \bm{\varepsilon} \right)  \, \text{d} V + \int_{V} G_c \gamma \left( \phi,\nabla \phi, \ell \right) \, \text{d} V  
\end{equation}

\noindent where $\gamma (\phi,\nabla \phi)$ is the so-called crack density functional and $g(\phi)$ is the degradation function. We choose to adopt the standard or \texttt{AT2} phase field formulation \cite{Bourdin2000}, and accordingly make the following constitutive choices, 
\begin{equation}
\label{eq:2.2.3}
g(\phi) = (1-\phi)^2 + \kappa \, ,
\end{equation}
\begin{equation}
\label{eq:2.2.4}
\gamma (\phi,\nabla, \ell \phi) = \frac{1}{2\ell}\phi^2 + \frac{\ell}{2} \lvert\nabla \phi\rvert^{2},
\end{equation}

\noindent where $\kappa$ is a small numerical parameter to retain residual stiffness when $\phi=1$, so as to avoid an ill-conditioned system of equations. Noting that $\bm{\sigma}=\partial_{\bm{\varepsilon}} \psi$, the strong form of the coupled deformation-fracture problem can be readily obtained by inserting (\ref{eq:2.2.3})-(\ref{eq:2.2.4}) into (\ref{eq:Potential}), taking the variation with respect to $\mathbf{u}$ and $\phi$, and applying Gauss theorem. However, such a formulation would predict cracking also under compressive stress states. Hence, we adopt the so-called volumetric-deviatoric split \cite{Amor2009} to decompose the elastic strain energy density into a tensile part,
\begin{equation}
\label{eq:2.2.6a}
\psi_{+}^e(\boldsymbol{\varepsilon}^e) = \frac{K}{2} \langle \text{tr} [\boldsymbol{\varepsilon}^e]\rangle_{+}^{2} + \mu [(\boldsymbol{\varepsilon}^{e})' : (\boldsymbol{\varepsilon}^{e})']
\end{equation}

\noindent and a compressive part,
\begin{equation}
\label{eq:2.2.6b}
\psi_{-}^e(\boldsymbol{\varepsilon}^e) = \frac{K}{2} \langle \text{tr}[\boldsymbol{\varepsilon}^e]\rangle_{-}^{2}
\end{equation}

\noindent Here, $K$ is the bulk modulus, $\text{tr}[\bullet]$ is the trace operator, $\langle \bullet \rangle = (\bullet \pm |\bullet|)/2$ and $(\boldsymbol{\varepsilon}^{e})' = \boldsymbol{\varepsilon}^{e} - \text{tr}[\boldsymbol{\varepsilon}^{e}]\mathbf{I}/3$. Furthermore, a history field $\mathcal{H}$ is defined to enforce damage irreversibility. Among the various choices available (see, e.g., \cite{Duda2015,Borden2016}), we choose to assume that fracture is driven by the energy stored in the system, in consistency with the energy balance above \cite{PTRSA2021}; accordingly: $\mathcal{H}=\text{max}_{\text{t}\in[0,\tau]} \, \psi_+^e(\boldsymbol{\varepsilon}^e, \text{t})$. The local governing equations then read,
\begin{equation}
\label{eq:2.2.8}
\boldsymbol{\nabla}\cdot \left\{ [(1-\phi)^2 + \kappa] \boldsymbol{\sigma}_0 \right\} = \mathbf{0} \quad \text{in} \ \Omega
\end{equation}
\begin{equation}
\label{eq:2.2.9}
{G}_{c}\bigg(\frac{1}{\ell}\phi - \ell \nabla^2 \phi\bigg) - 2(1-\phi)\mathcal{H} = 0 \quad \text{in} \ \Omega
\end{equation}

\noindent where $\boldsymbol{\sigma}_0$ is the undamaged Cauchy stress tensor. As can be inferred by inspecting (\ref{eq:2.2.8})-(\ref{eq:2.2.9}), a so-called hybrid approach is used, where the strain energy split is considered only in the phase field evolution equation \cite{Ambati2015}.

\subsection{Hydrogen-sensitive interface formulation for grain boundaries}
\label{SubSec:Int}

In this modelling framework, the polycrystalline nature of the material is resolved and the decohesion of grain boundaries is explicitly captured by means of a cohesive zone model. Specifically, a traction-separation law is adopted that assumes a tension cut-off relation \cite{Paggi2017}. This interface formulation assumes a linear and reversible (elastic) evolution until the critical traction is reached. The damage criterion relates the normal $t_n$ and tangential $t_t$ tractions with their critical counterparts as follows,
\begin{equation}
\label{eq:2.4.1}
\bigg(\frac{t_n}{t_{nc}}\bigg)^2 + \bigg(\frac{t_t}{t_{tc}}\bigg)^2 = 1
\end{equation}

Accordingly, a critical normal $\delta_{nc}$ and shear $\delta_{tc}$ separations can be defined, which leads to the following definitions of fracture energy for Mode I and Mode II fractures
\begin{equation}
\label{eq:2.4.2}
\gamma_{IC}=\frac{1}{2}t_{nc}\delta_{nc}, \quad \gamma_{IIC}=\frac{1}{2}t_{tc}\delta_{tc}
\end{equation}

As sketched in Fig. \ref{fig:Sketch}, the role of hydrogen in weakening the grain boundaries is accounted for by degrading the interface fracture energy. Thus, the focus here is on materials that exhibit hydrogen assisted intergranular fracture, such a Ni and Ni alloys \cite{Harris2018}. Atomistic calculations have shown a linear relationship between the surface (or fracture) energy and the hydrogen coverage (see, e.g., \cite{Jiang2004a,Alvaro2015}). Accordingly, and following Mart\'{\i}nez-Pa\~neda \textit{et al.} \cite{CMAME2018}, we define the following relationship between the fracture energy and the hydrogen coverage $\theta_H$,
\begin{equation}
\label{eq:2.4.3}
{\gamma_c}( \theta_H) = {\gamma_{C,0}}(1 - \chi\theta_H)
\end{equation}

\noindent where $\gamma_{C,0}$ is the fracture energy in the absence of hydrogen and $\chi$ is the so-called hydrogen damage coefficient \cite{CMAME2018}. The same expression is employed for $\gamma_{IC}$ and $\gamma_{IIC}$. Finally, the hydrogen coverage is estimated from the hydrogen concentration by means of the Langmuir-McLean isotherm:
\begin{equation}
\label{eq:2.4.4}
\theta_H = \frac{C}{C+\exp \bigg(\frac{-\Delta g_{b}^{0}}{RT}\bigg)}
\end{equation}

\noindent where $\Delta g_{b}^{0}$ is the Gibbs free energy difference between the interface and the surrounding material, also referred to as the segregation energy. Unless indicated otherwise, a value of $\Delta g_{b}^{0}= \ 30\text{kJ}/\text{mol}$ is here employed, based on the spectrum of experimental data available for the trapping energy at grain boundaries \cite{CMAME2018,Serebrinsky2004}. 

\subsection{Numerical implementation}

\begin{figure}[t]
	\centering
	\includegraphics[width=17cm]{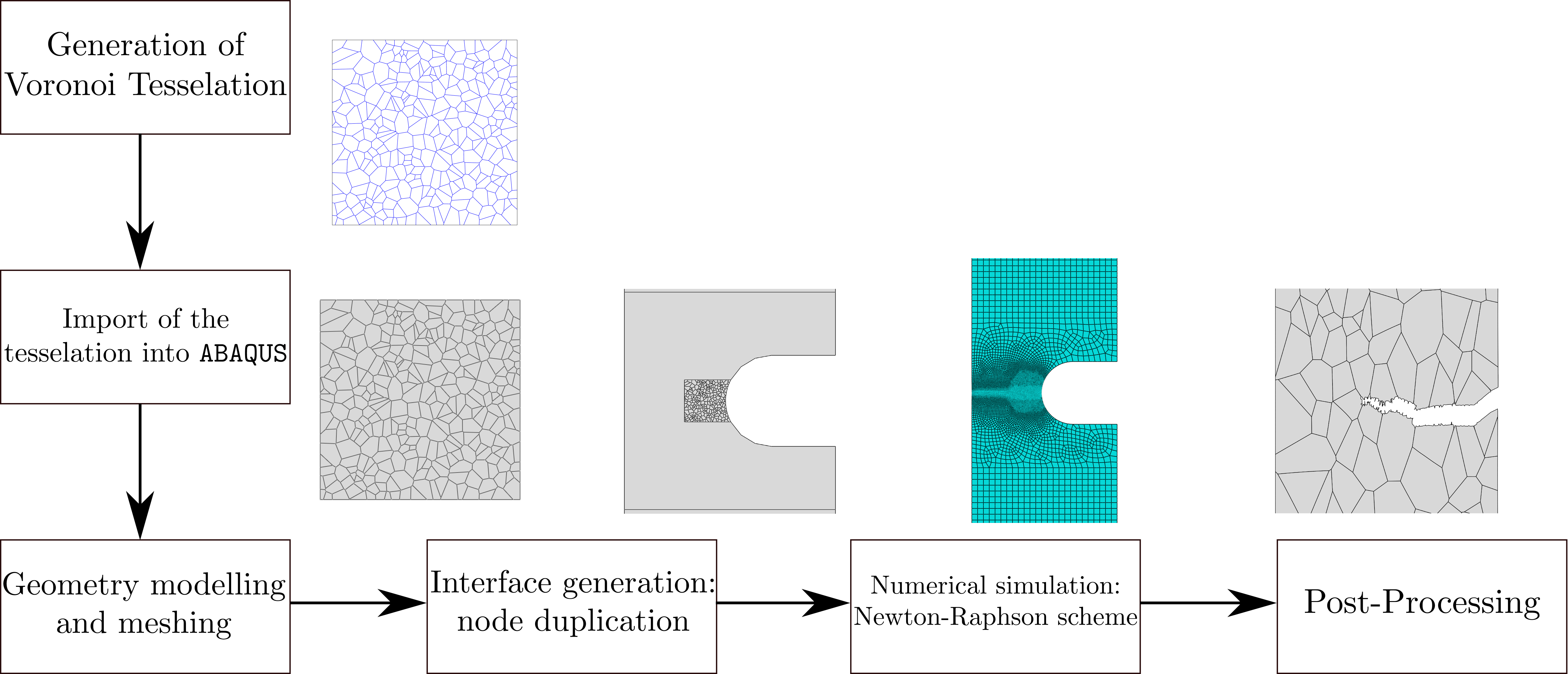}
	\caption{Flowchart that summarises the methodology adopted in the numerical deformation-diffusion-fracture analysis of polycrystalline solids exposed to hydrogen.} 
	\label{fig:flowchart}
\end{figure}

The theoretical model presented in Sections \ref{SubSec:Introduction} to \ref{SubSec:Int} is numerically implemented by means of the finite element method. Specifically, the model is implemented in the commercial finite element package \texttt{ABAQUS}/Standard \textit{via} a user element (\texttt{UEL}) subroutine. In addition, the \texttt{Abaqus2Matlab} software \cite{AES2017} is used to generate the input files and the \texttt{MATLAB} supplementary codes given in Ref. \cite{Paggi2018} are used to generate the microstructure. Fig. \ref{fig:flowchart} provides a flowchart of the steps followed in the definition and analysis of the microstructure sensitive boundary value problems investigated. Specifically, the microstructure is generated by using a Voronoi-based tessellation algorithm, programmed in \texttt{MATLAB} \cite{Paggi2018}, and this step is followed by the introduction of the resulting microstructure into the \texttt{ABAQUS} input file using a \texttt{Python} script. The coupled deformation-diffusion-fracture system is solved in a staggered fashion, with every sub-problem being solved by means of a backward Euler solution scheme. Typical calculation times are of a few hours.

\section{Results}
\label{Sec:FEMresults}

The potential of the theoretical and computational framework presented in Section \ref{Sec:Form} shall be demonstrated through representative case studies. First, in Section \ref{SubSec:Bench}, a benchmark example is analysed, whereby cracking is predicted in a single edged notched tension specimen containing 50 grains and exposed to various hydrogen-containing environments. Second, we simulate recent slow strain rate tests (SSRTs) on different Monel K-500 lots in Section \ref{SubSec:SSRT}, so as to assess the ability of the model in providing a quantitative agreement with experiments and shedding light into the suitability of SSRTs to measure hydrogen susceptibility. Finally, in Section \ref{SubSec:Temp}, the model is used to simulate, for the first time, the seminal experiments by Harris \textit{et al.} \cite{Harris2018} on pure Ni under cryogenic and ambient temperature conditions. 

\subsection{Benchmark: fracture of a 50-grain SENT plate}
\label{SubSec:Bench}

\begin{figure}[H]
\centering

\tabskip=0pt
\valign{#\cr
  \hbox{%
    \begin{subfigure}[b]{.35\textwidth}
    \centering
    \vspace*{1cm}
    \includegraphics[width=\textwidth]{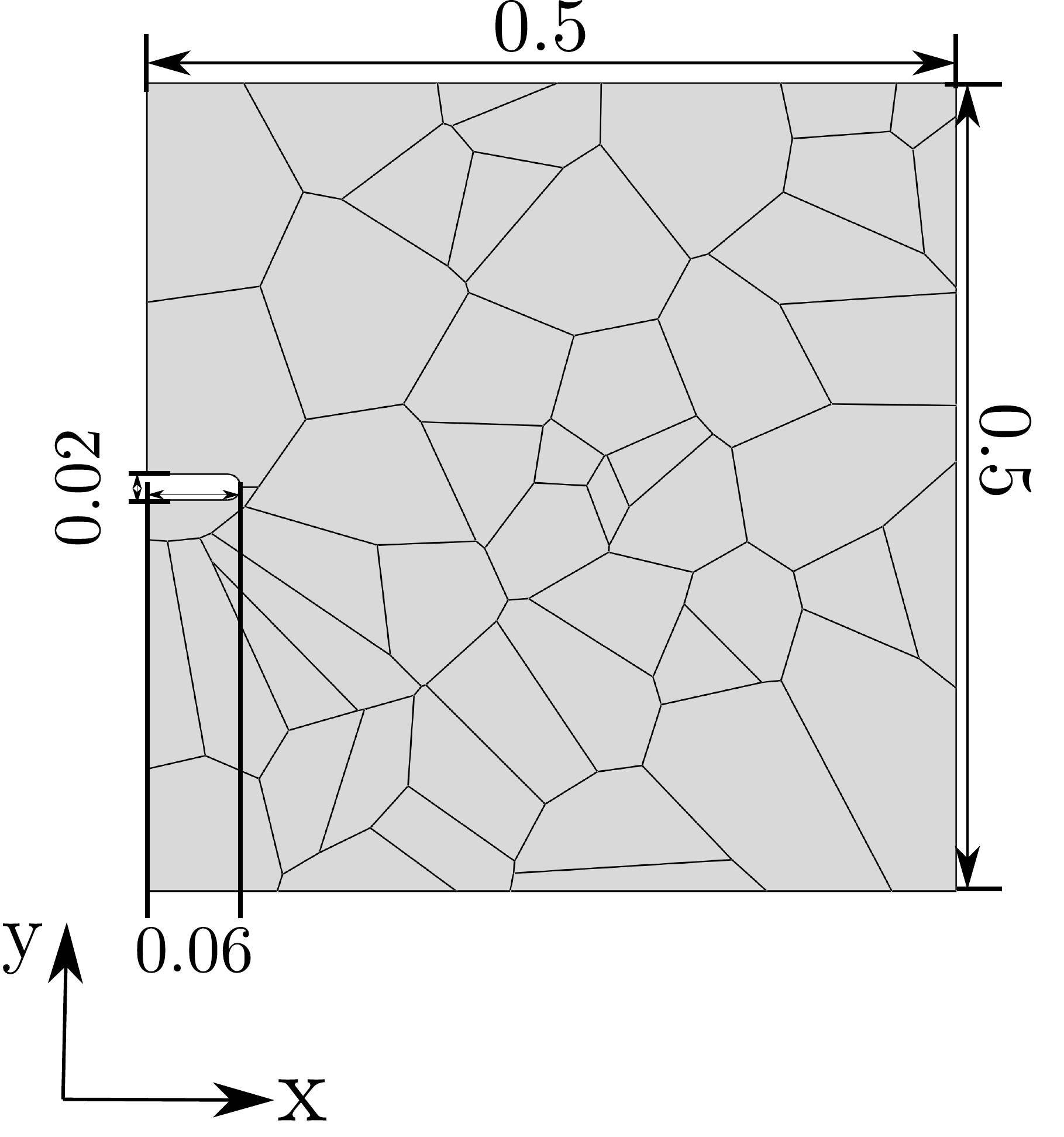}
    \caption{Geometry}
    \end{subfigure}%
  }\cr
  \noalign{\hfill}
  \hbox{%
    \begin{subfigure}{.20\textwidth}
    \centering
    \includegraphics[width=\textwidth]{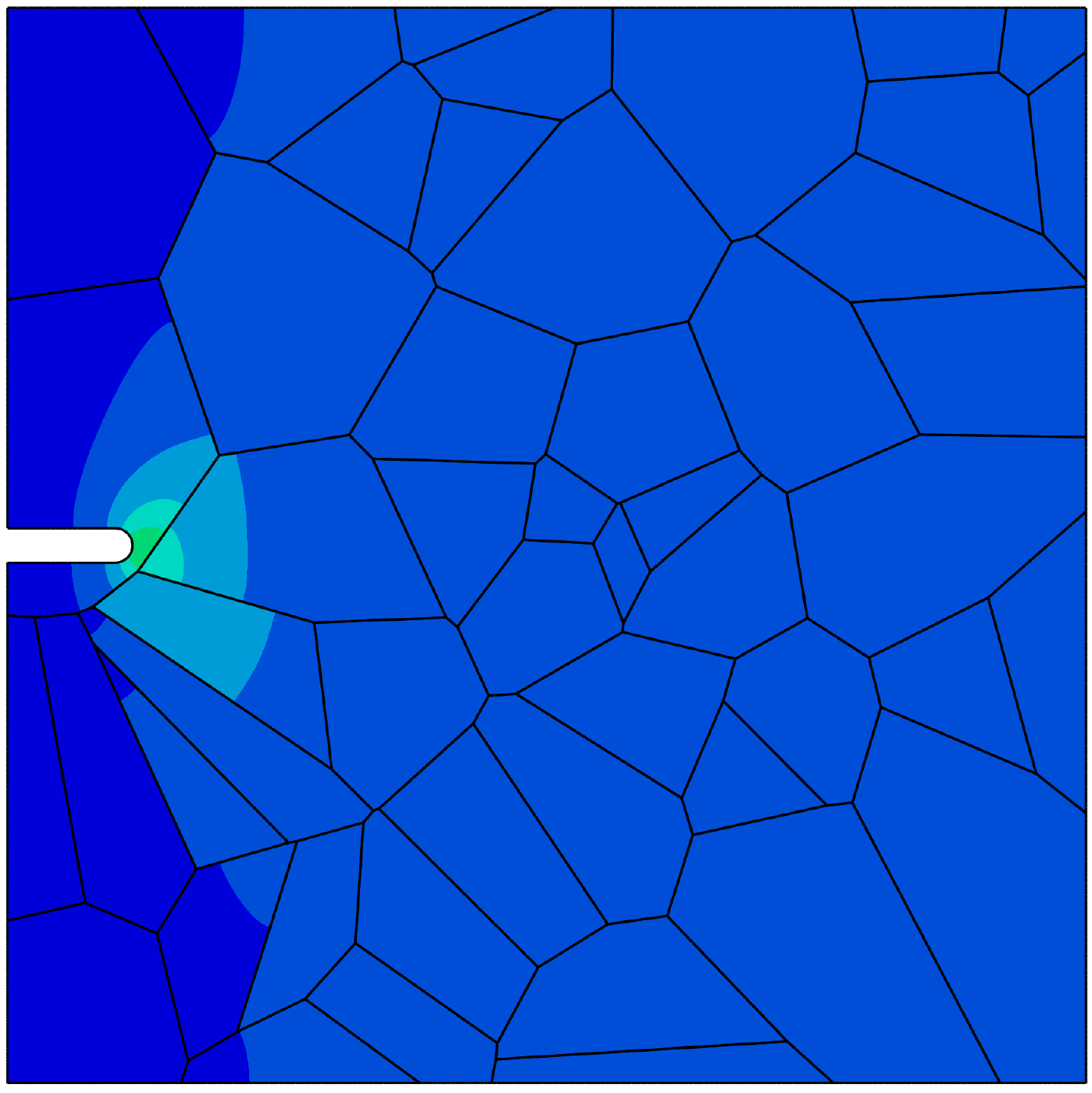}
    \caption{$u=5.96\times10^{-4} \ \text{mm}$}
    \end{subfigure}%
    \begin{subfigure}{.20\textwidth}
    \centering
    \includegraphics[width=\textwidth]{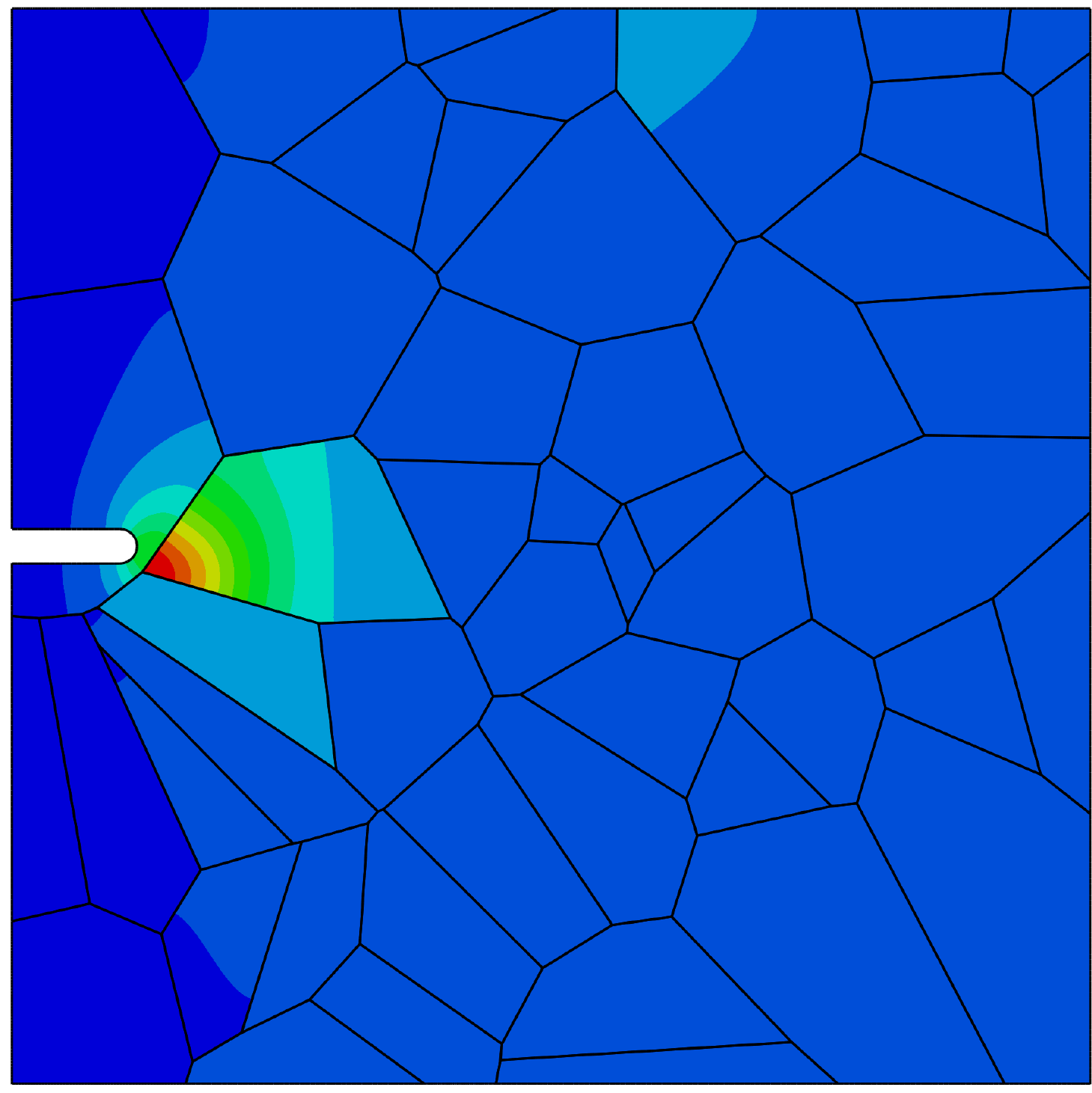}
    \caption{$u=6.68\times10^{-4} \ \text{mm}$}
    \end{subfigure}%
    \begin{subfigure}{.20\textwidth}
    \centering
    \includegraphics[width=\textwidth]{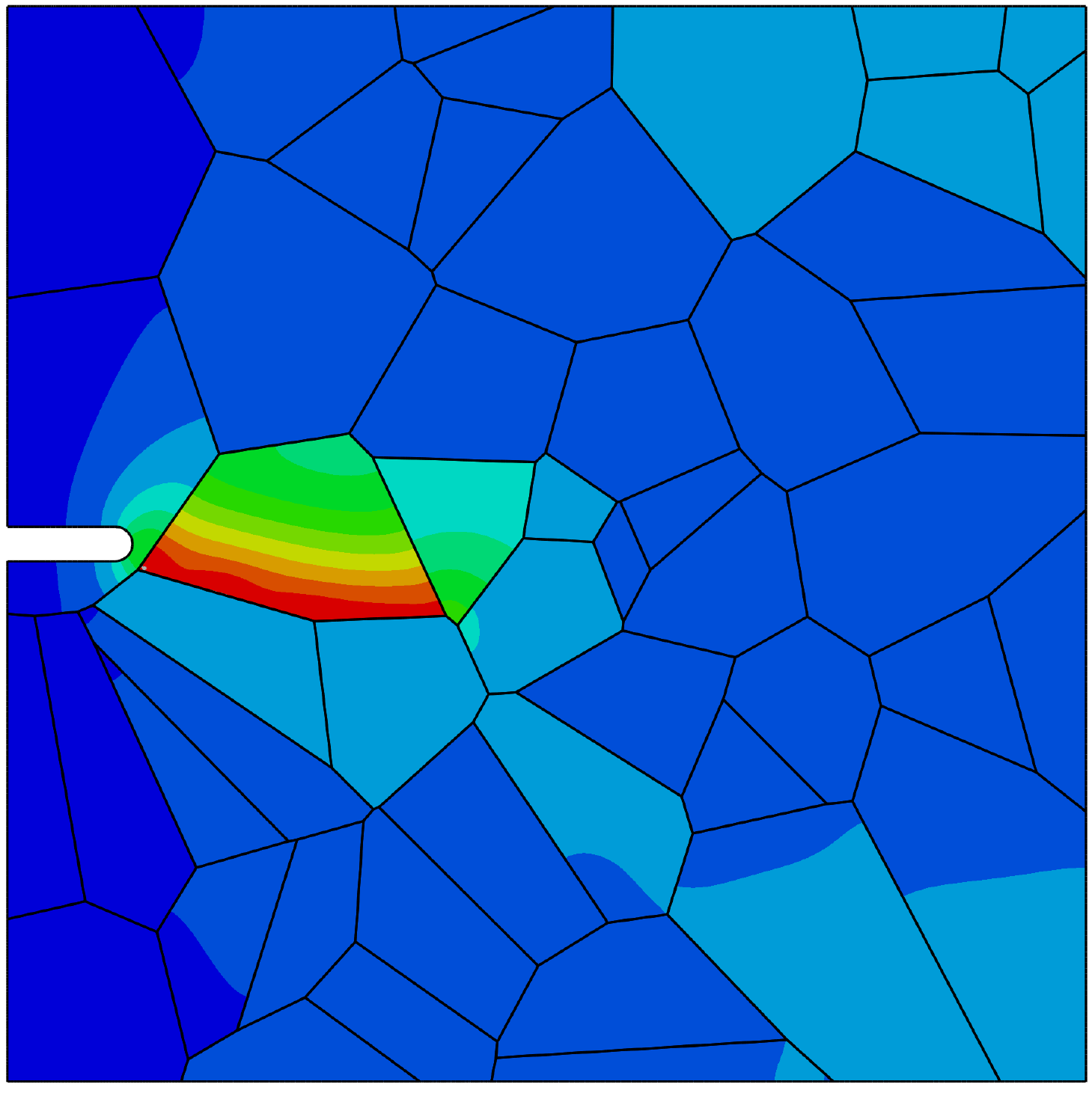}
    \caption{$u=6.97\times10^{-4} \ \text{mm}$}
    \end{subfigure}%
  }
  \vfill
  \hbox{%
    \begin{subfigure}{.20\textwidth}
    \centering
    \includegraphics[width=\textwidth]{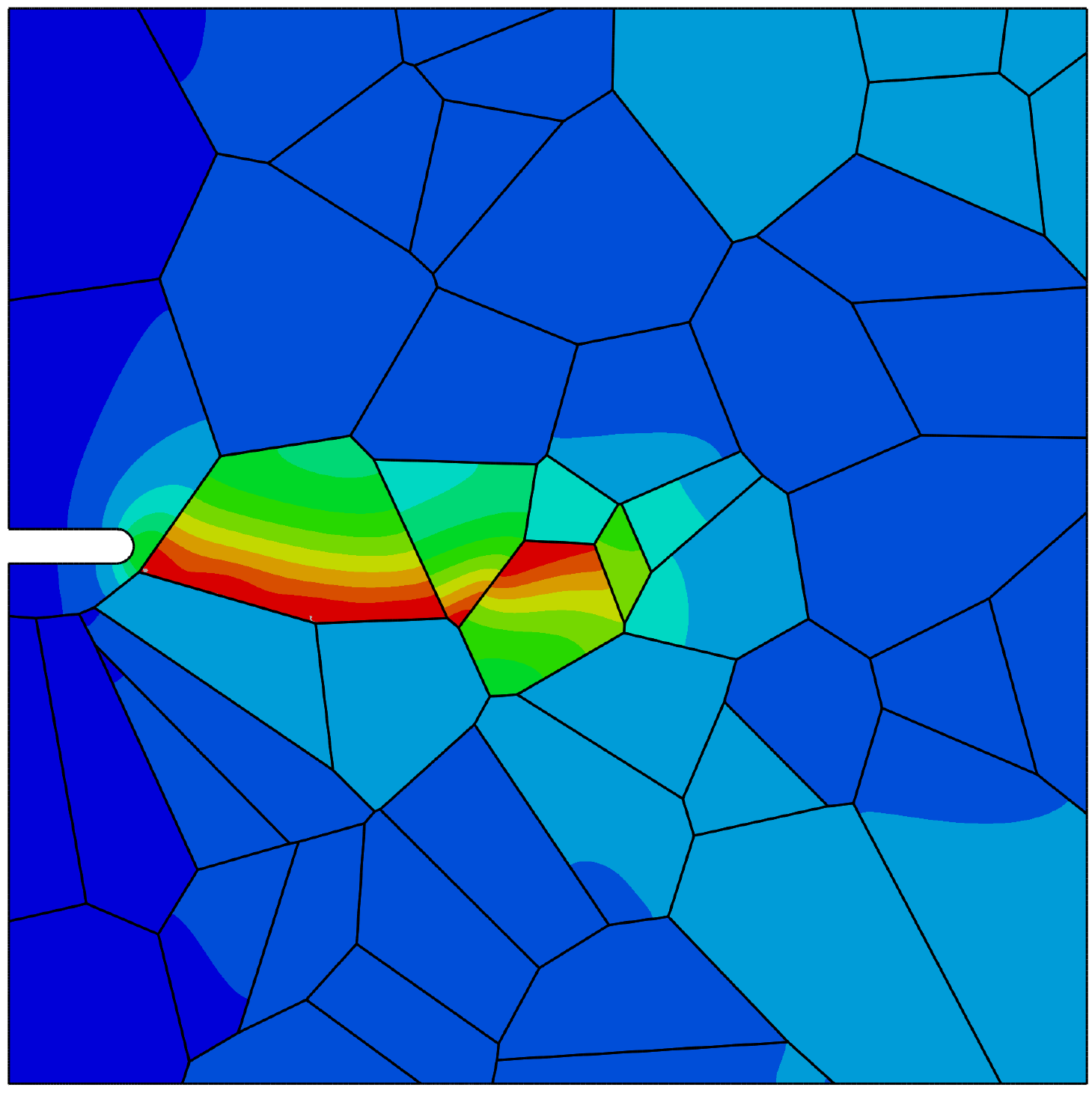}
    \caption{$u=6.97\times10^{-4} \ \text{mm}$}
    \end{subfigure}%
    \begin{subfigure}{.20\textwidth}
    \centering
    \includegraphics[width=\textwidth]{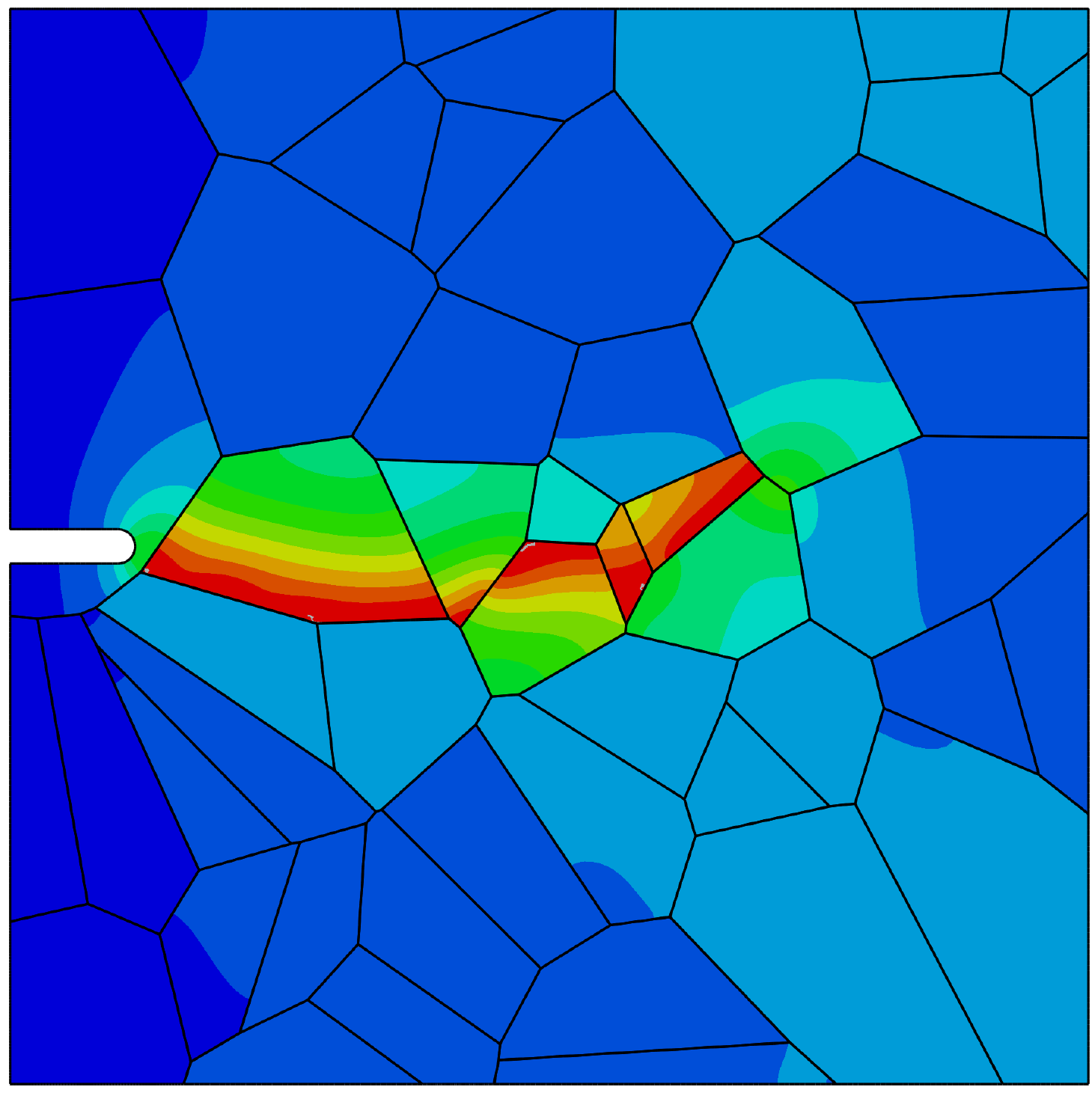}
    \caption{$u=6.97\times10^{-4} \ \text{mm}$}
    \end{subfigure}%
    \begin{subfigure}{.20\textwidth}
    \centering
    \includegraphics[width=\textwidth]{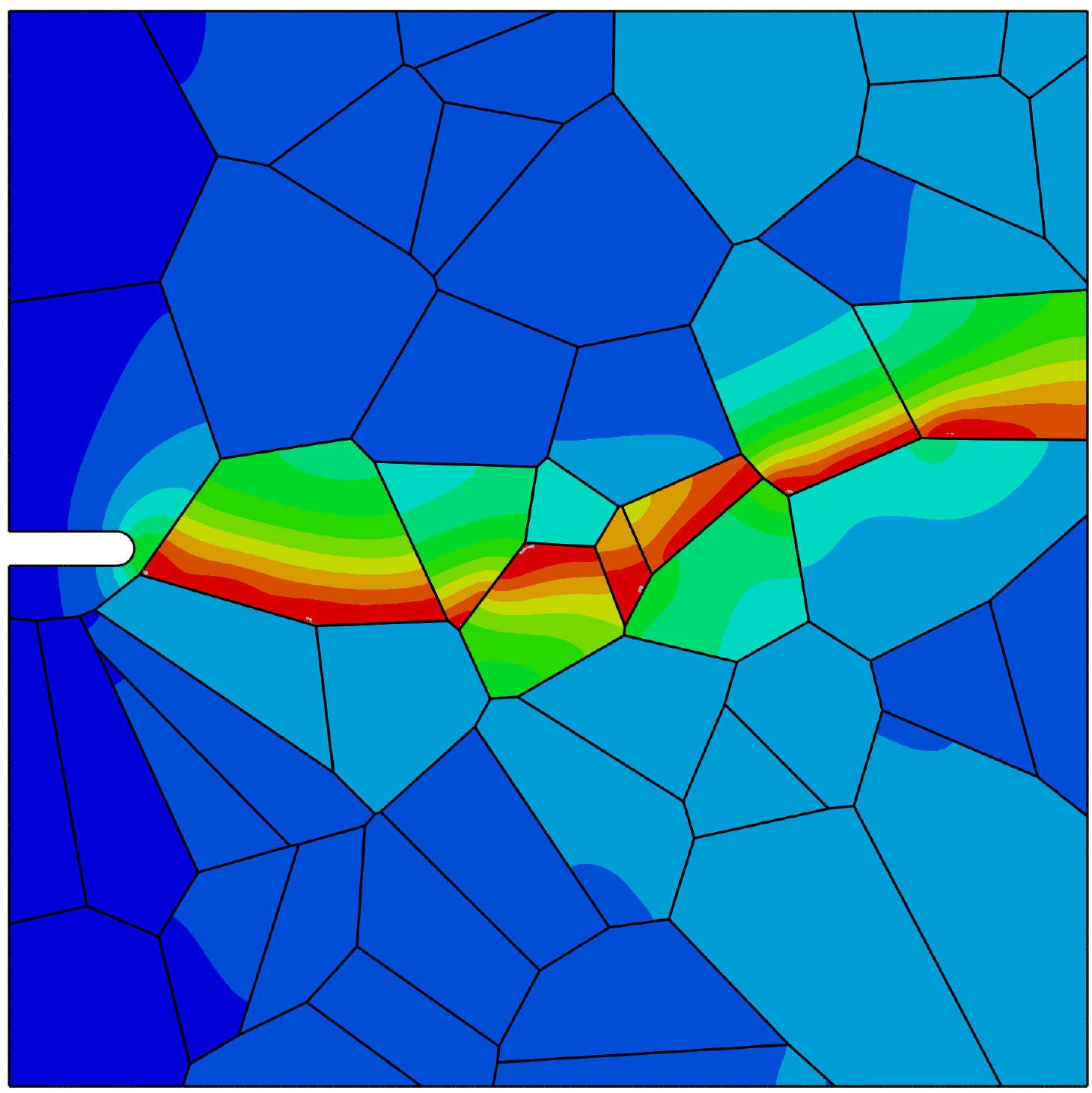}
    \caption{$u=6.97\times10^{-4} \ \text{mm}$}
    \end{subfigure}%
  }\cr
}

\caption{Single edge notched tension (SENT) specimen: (a) geometry, with dimensions in mm, and (b)-(f) phase field contours as a function of the applied displacement. Results obtained for the case of $C_{\text{env}}=0$ ppm wt, with cracking occurring in a transgranular manner due to the absence of hydrogen. The phase field contours are shown in a sequential manner, with the last snapshots occurring closely in time due to the unstable nature of the cracking process at that stage (see Fig. \ref{fig:50G-RFU}).}
\label{fig:SENT-TG}
\end{figure}

Crack nucleation and growth in a square plate microstructure of 50 grains is investigated. The plate, with dimensions given in Fig. \ref{fig:SENT-TG}a, contains a small notch and is subjected to uniaxial loading; a testing configuration often referred to as single edge notched tension (SENT) specimen. The load is applied by prescribing the vertical displacement at the upper edge, at a rate of $\dot{u}_y=10^{-10}$ mm/s, while the vertical displacement is constrained at the bottom edge. To prevent rigid body motion, the horizontal displacement is constrained at the bottom-right corner. The sample is exposed to hydrogen on its left side, where the notch is located. No pre-charging time in considered, with both hydrogen and mechanical charging starting simultaneously. The magnitude of the environmental hydrogen concentration $C_{\text{env}}$ is varied between 0 and 0.9 ppm wt with the aim of capturing a reduced critical load with increasing hydrogen content, and a transition from ductile (bulk) fracture to intergranular cracking. The material properties assumed for the bulk and the interface are given in Table \ref{tab:MatPropsBenchmark}. The sample is discretised using a total of 48,554 quadratic quadrilateral elements.

\begin{table}[H]
\centering
\caption{Material properties of the bulk and the interface adopted in the SENT benchmark case study.}
\label{tab:MatPropsBenchmark}
\begin{tabular}{@{}cccccccl@{}}
\toprule
\multicolumn{7}{c}{\textbf{Bulk properties}}             
\\ \midrule
$E \ (\text{GPa})$              & $\nu$              & $\ell \  (\text{mm})$ & $G_c \ (\text{kJ}/\text{m}^2)$ & $D \ (\text{mm}^2/\text{s})$ & $ \sigma_y \ \mathbf{ (\text{MPa})}$ & n          \\
\midrule
185                                             & 0.3                             & 0.025                                & 0.05                                                                   & $1.3 \times 10^{-8}$                                 & 794.3                                                    & 0.064               \\ \midrule
\multicolumn{7}{c}{\textbf{Interface properties}}                                                                                                                                                                                                                                                                                         \\ \midrule
\multicolumn{2}{c}{$k_n,k_t \ (\text{MPa}/\text{mm})$} & \multicolumn{2}{c}{$t_{nc,0},t_{tc,0} \ (\text{MPa})$}                                      & \multicolumn{2}{c}{$\gamma_{IC,0},\gamma_{IIC,0} \ (\text{kJ}/\text{m}^2)$} & $\chi$ \\ \midrule
\multicolumn{2}{c}{$2\times10^8$}                                                 & \multicolumn{2}{c}{$2.25\times10^3$}                                                                          & \multicolumn{2}{c}{$1.27\times10^{-2}$}                                                                         & 0.86                \\ \bottomrule
\end{tabular}
\end{table}

The results obtained for the case of $C_{\text{env}}=0$ ppm wt are shown in Figs. (\ref{fig:SENT-TG}b-\ref{fig:SENT-TG}g), in terms of the phase field contours. In the absence of hydrogen, the crack nucleates inside of the grain, in the vicinity of the notch tip, and propagates in a transgranular manner. The damage appears to accumulate within the grain region closer to its boundary, as a stress mismatch takes place between adjacent grains due to differences between the stiffness of the bulk and that of the interface. 

\begin{figure}[H]
\centering
\subfloat[][$u=2.75\times10^{-4} \ \text{mm}$\label{fig:3.2.3a}]
{\includegraphics[width=0.24\linewidth]{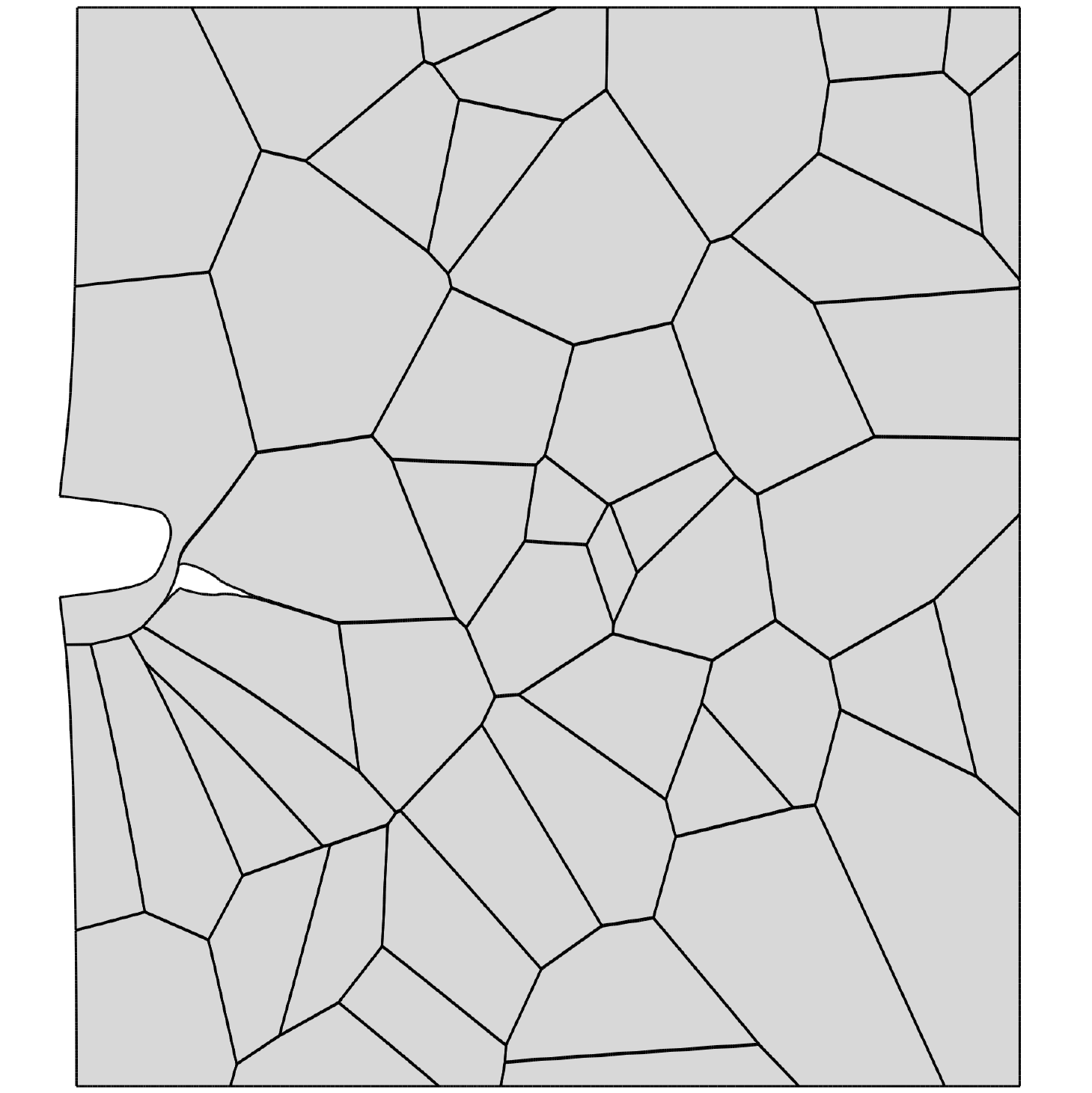}}
\subfloat[][$u=2.75\times10^{-4} \ \text{mm}$\label{fig:3.2.3b}]
{\includegraphics[width=0.24\linewidth]{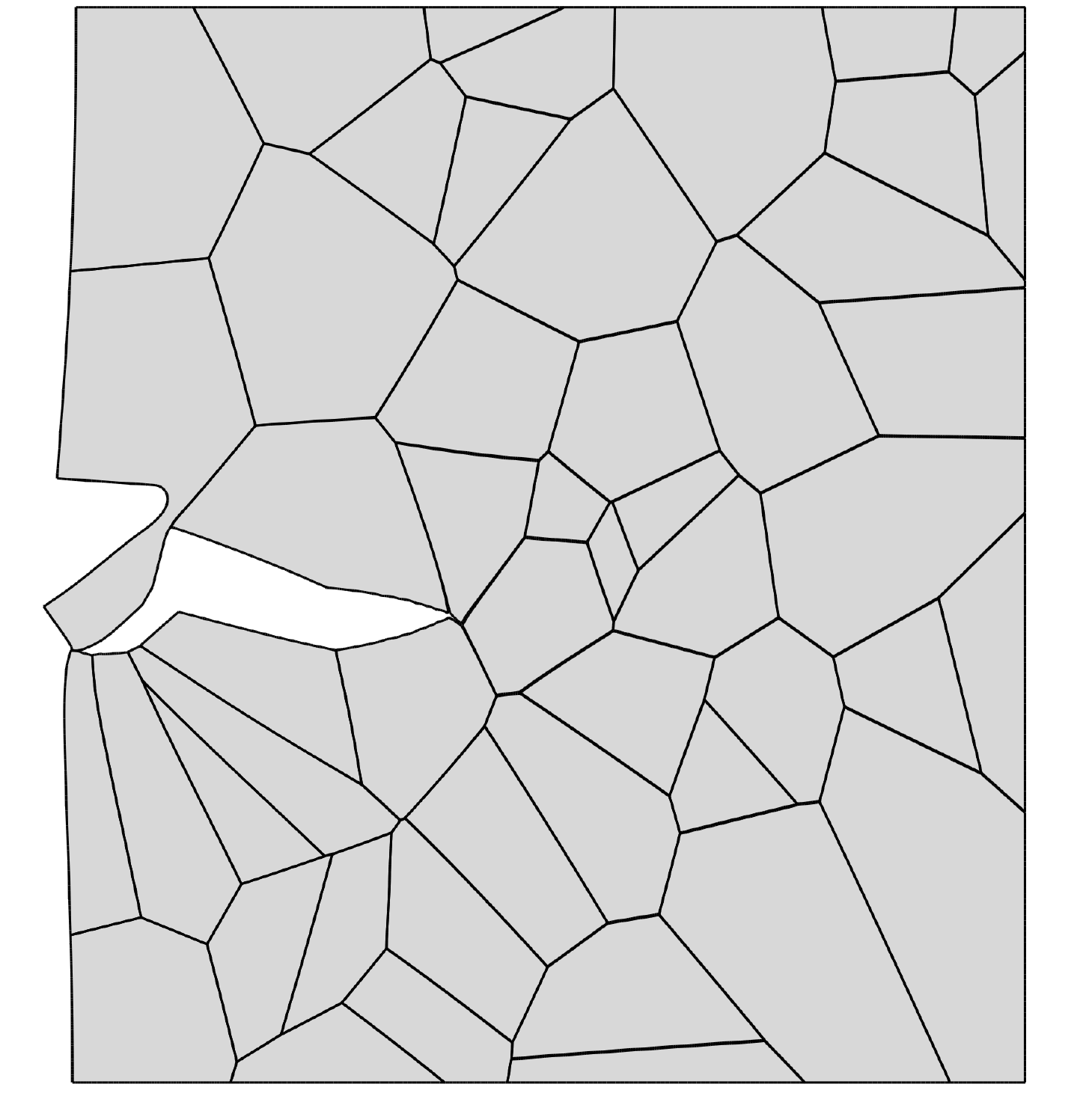}}
\subfloat[][$u=2.75\times10^{-4} \ \text{mm}$\label{fig:3.2.3c}]
{\includegraphics[width=0.24\linewidth]{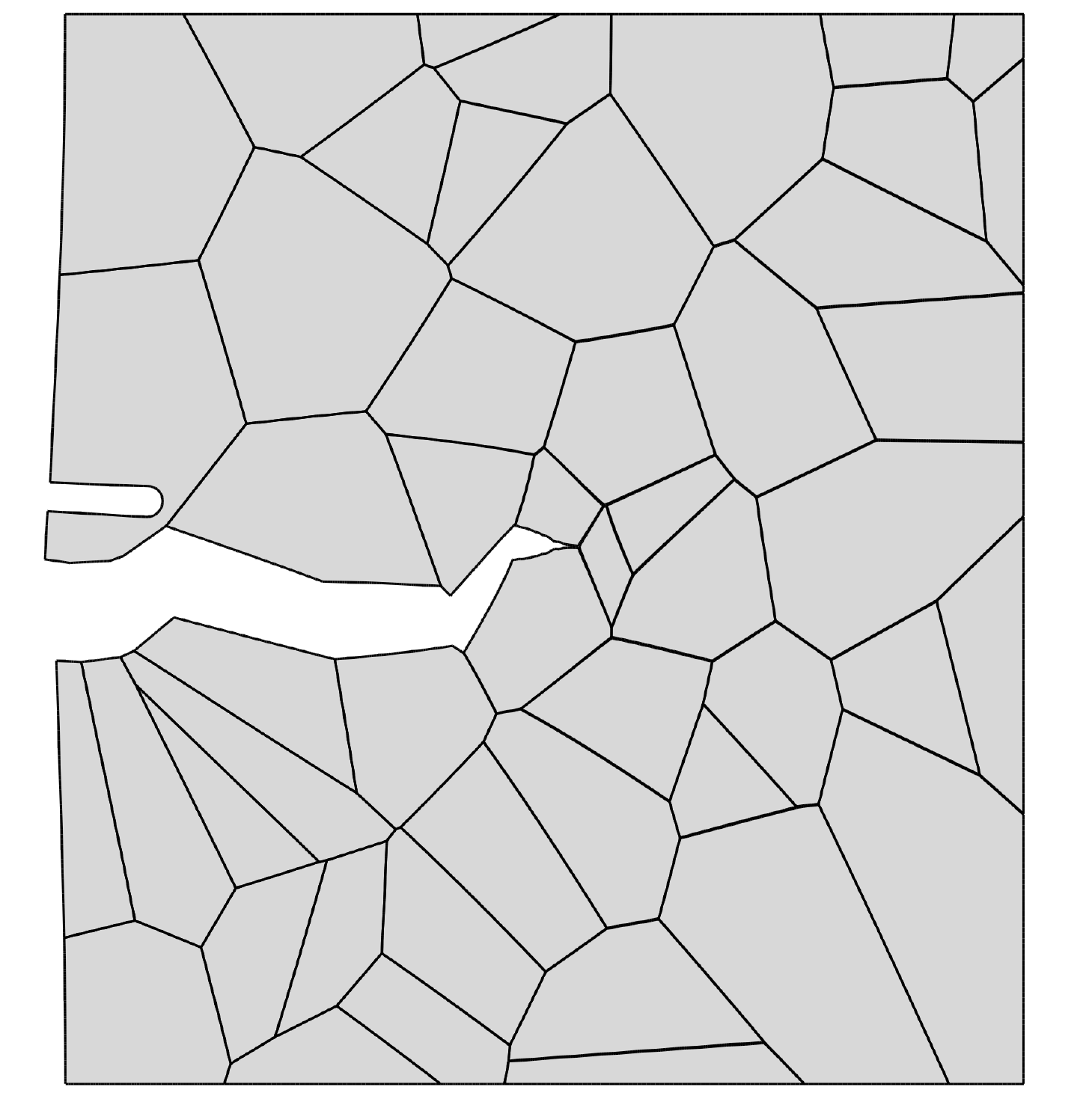}}\\
\subfloat[][$u=2.75\times10^{-4} \ \text{mm}$\label{fig:3.2.3d}]
{\includegraphics[width=0.24\linewidth]{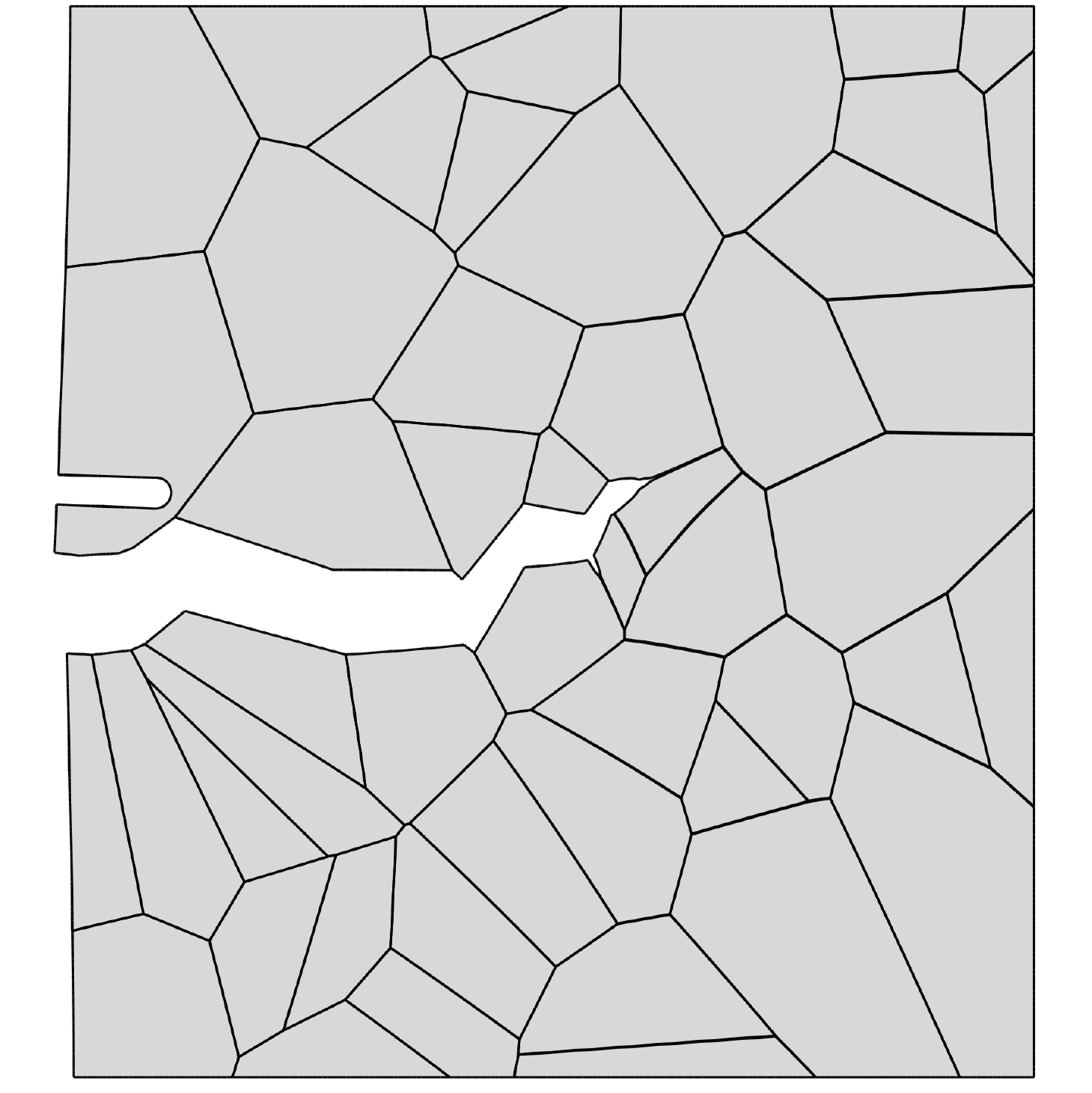}}
\subfloat[][$u=2.75\times10^{-4} \ \text{mm}$\label{fig:3.2.3e}]
{\includegraphics[width=0.24\linewidth]{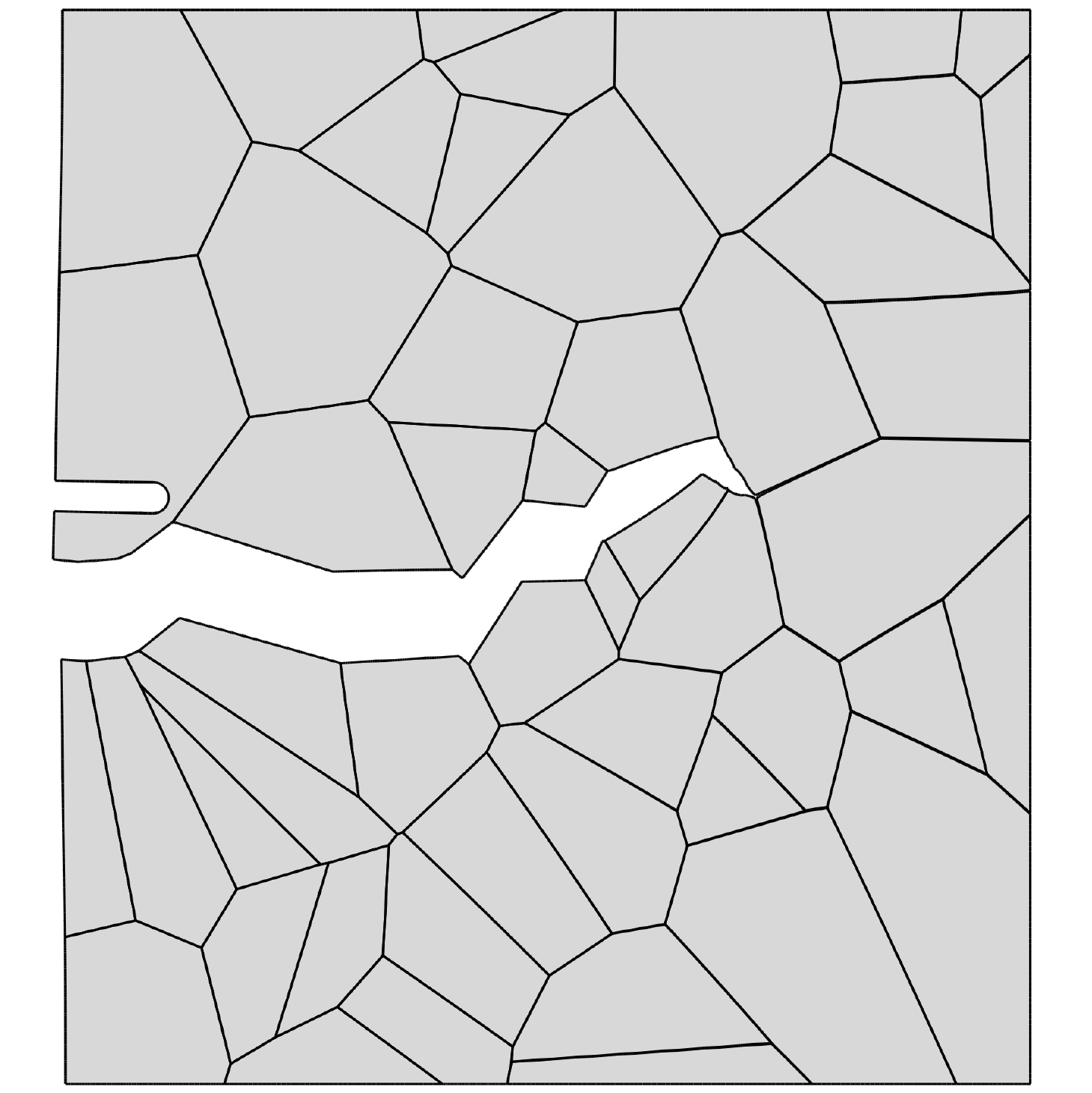}}
\subfloat[][$u=2.75\times10^{-4} \ \text{mm}$\label{fig:3.2.3f}]
{\includegraphics[width=0.24\linewidth]{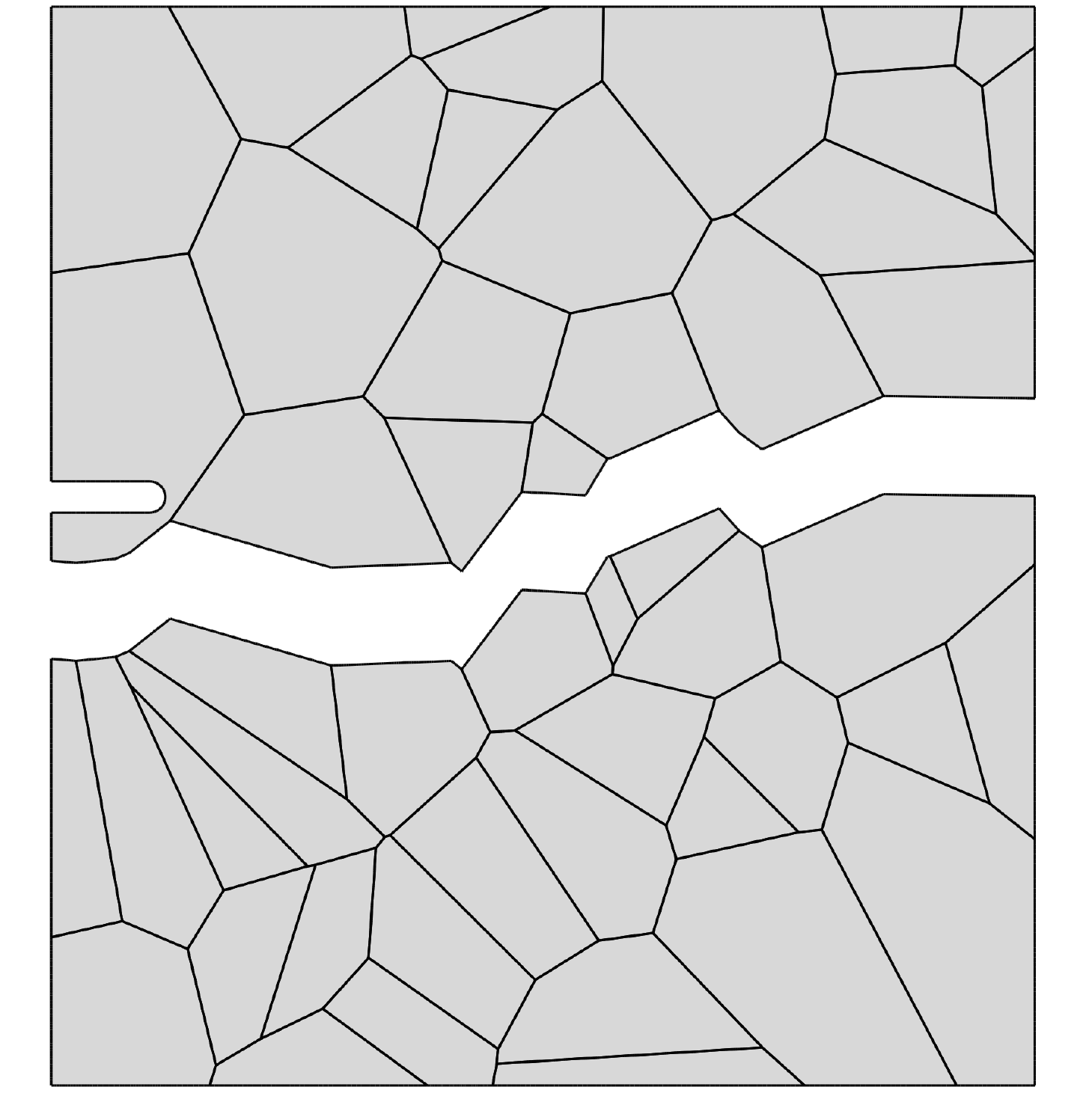}}\\
\caption{Cracking evolution for the SENT benchmark case study as a function of the applied displacement. Results obtained for the case of $C_{\text{env}}=0.25$ ppm wt, with cracking occurring in an intergranular manner as a result of the weakening of the grain boundaries due to the presence of hydrogen. 
The cracking contours are shown in a sequential manner. Since cracking occurs rapidly (see Fig. \ref{fig:50G-RFU}), all the snapshots correspond to approximately the same applied displacement.}
\label{fig:SENT-IG}
\end{figure}

Contrarily, when the sample is exposed to hydrogen, cracking initiates along the grain boundaries and propagates in an intergranular manner. This is shown in Fig. \ref{fig:SENT-IG}, where the deformation and separation of the grains is shown as a function of the applied displacement for the case of $C_{\text{env}}=0.25$ ppm wt. The crack nucleates close to the notch, where both hydrogen content and tensile stresses are large (see Fig. \ref{fig:SENT-IG}a). As the remote load increases, the crack spreads to neighboring grain boundaries (Figs. \ref{fig:SENT-IG}b-e) and eventually leads to the complete failure of the specimen (Fig. \ref{fig:SENT-IG}f). Thus, by comparing Figs. \ref{fig:SENT-TG} and \ref{fig:SENT-IG}, one can see that the model captures the ductile (transgranular) to brittle (intergranular) transition typically observed in the presence of hydrogen. In the absence of hydrogen, the strength of the grain boundaries is sufficiently large for failure to be driven by other mechanisms, encapsulated here in the phase field fracture model. However, when there is sufficient hydrogen in the sample, the grain boundaries experience a drop in strength that leads to an early nucleation and growth of intergranular cracks. 

\begin{figure}[H]
\centering
\includegraphics[width=0.99\textwidth]{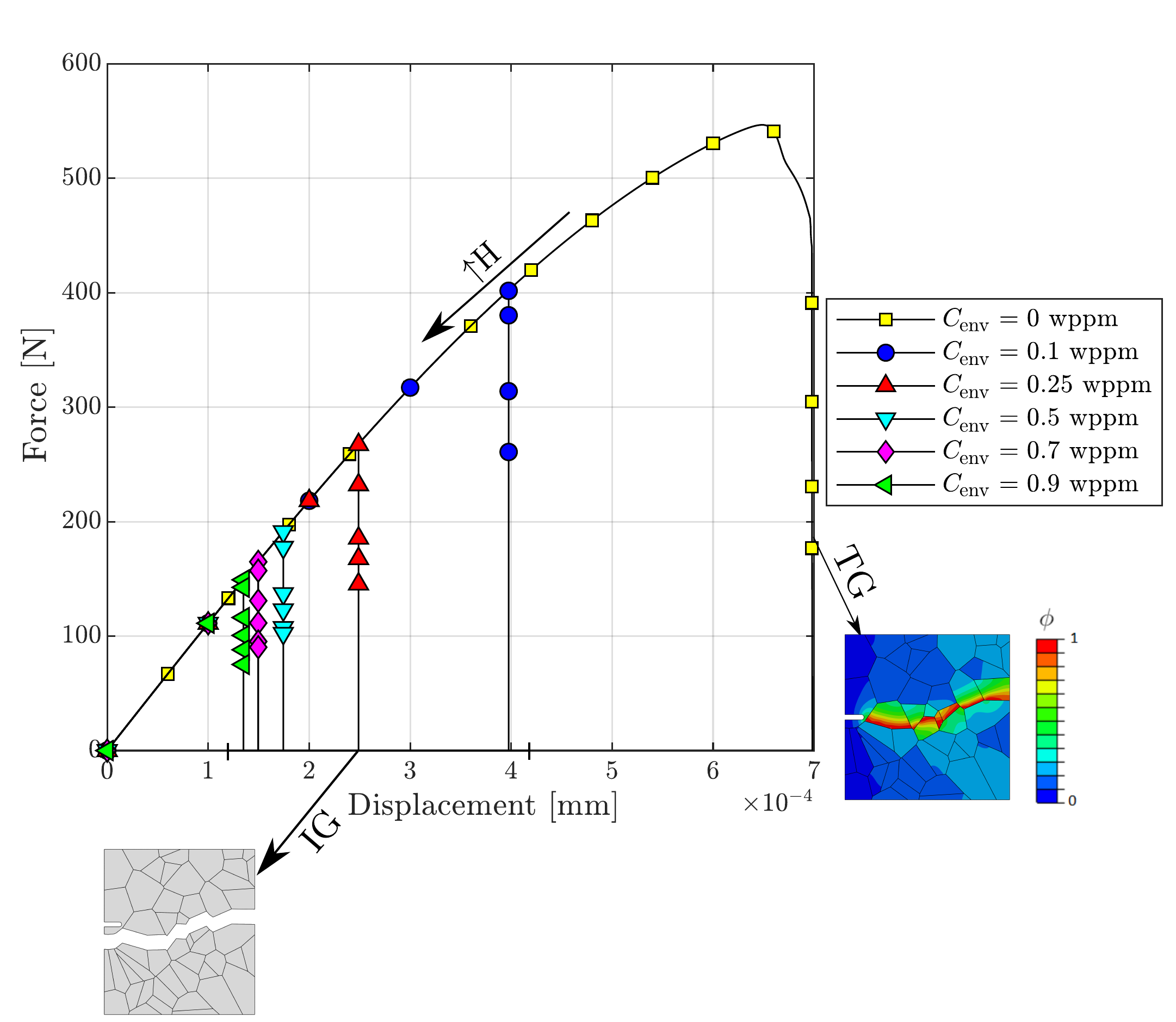}%
\caption{Predicted force versus displacement curves for the 50-grain SENT benchmark. The results are presented as a function of the environmental hydrogen concentration $C_{\text{env}}$, and images are embedded to showcase the different intergranular (IG) and transgranular (TG) cracking patterns observed.}
\label{fig:50G-RFU}
\end{figure}

Finally, the resulting force versus displacement curves are shown in Fig. \ref{fig:50G-RFU}. The result shows that the model not only captures the shift in cracking patterns but can also predict the progressive degradation of fracture properties with increasing hydrogen content. Furthermore, the results also showcase the robustness of the numerical model, as the entire fracture process can be captured, until the complete rupture and loss of load carrying capacity. 

\subsection{Virtual slow strain rate testing on Monel K-500 samples}
\label{SubSec:SSRT}

\begin{figure}[H]
\centering
\subfloat[][\label{fig:3.3.1a}]
{\includegraphics[width=0.2\linewidth]{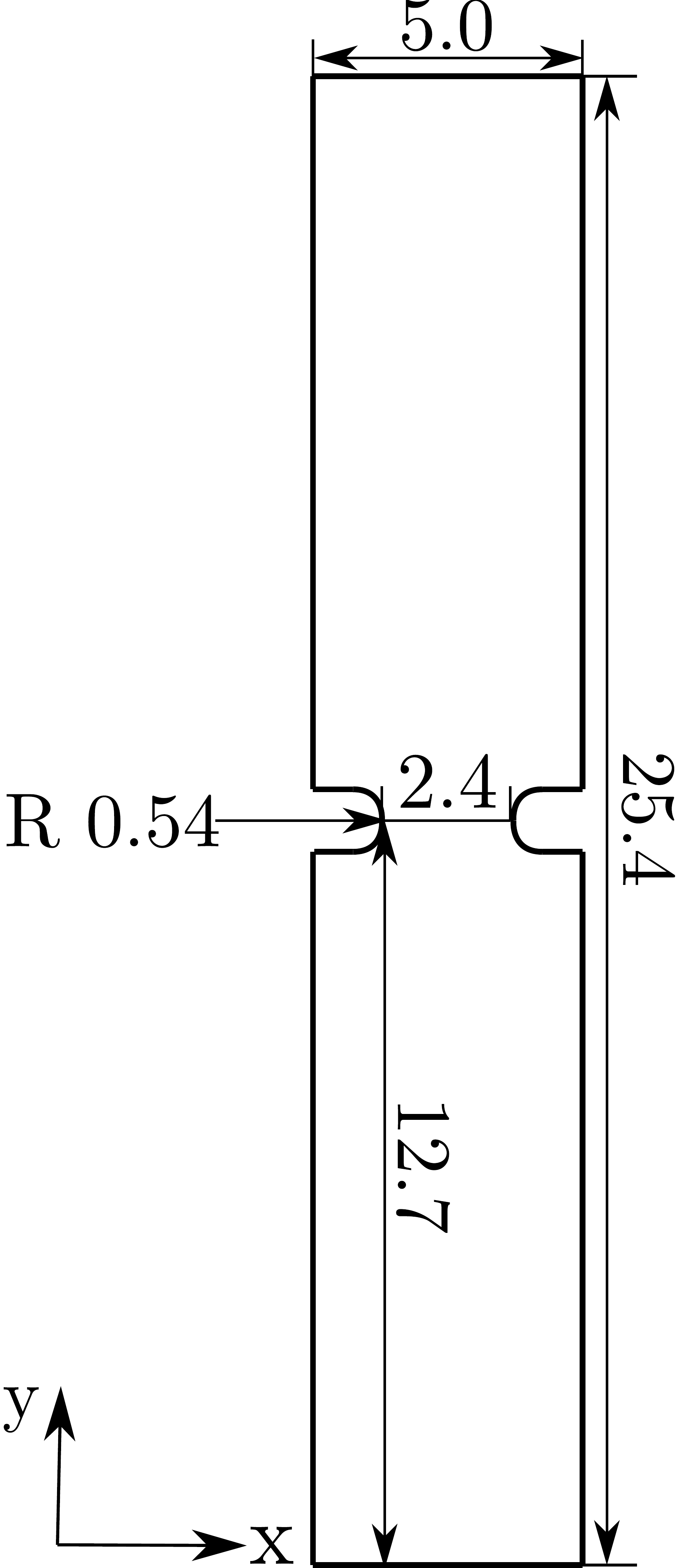}}
\hspace{0.2cm}
\subfloat[][\label{fig:3.3.1b}]
{\includegraphics[width=0.6\linewidth]{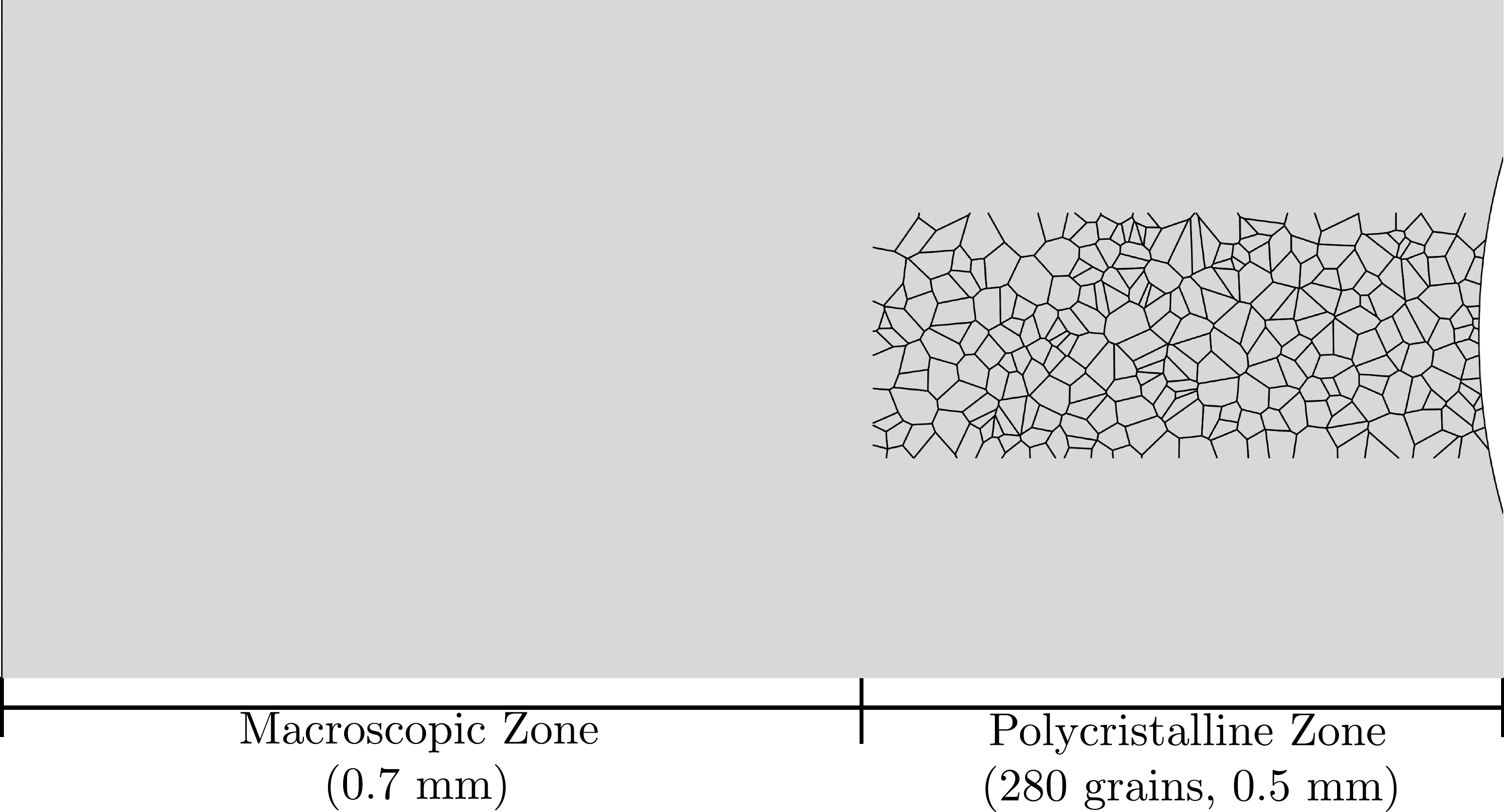}}
\caption{Virtual SSRT experiments on Monel K-500: (a) geometry of the specimens, with dimensions in mm, and (b) augmented view of the region ahead of the notch, showcasing the division between macroscopic and microscopic regions.}
\label{fig:MonelK500-Geom}
\end{figure}

Recent slow strain rate tests (SSRTs) on a Ni–Cu superalloy (Monel K-500) have revealed significant intergranular cracking depths, much larger than those expected from diffusion calculations for relevant exposure times \citep{CS2020}. Hence, these experiments suggest that crack initiation is likely to take place much before final failure, allowing for the hydrogen-containing solution to reach significant depths by following the crack. Early sub-critical crack growth would imply the need for a fracture analysis, compromising the suitability of the SSRT experiment and its metrics (e.g., time to failure) in assessing hydrogen embrittlement susceptibility. The micromechanical formulation presented can be used to gain complementary insight into these paradigmatic experiments and the early cracking hypothesis. Mimicking the testing conditions, notched cylindrical specimens with the dimensions shown in Fig. \ref{fig:MonelK500-Geom}a are considered in our simulations. The samples are subjected to uniaxial loading and thus one can take advantage of axially symmetric conditions. Accordingly, the sample is discretised using axisymmetrical finite elements; a total of 64,835 quadratic elements are employed. A microscopic region is introduced near the notch, see Fig. \ref{fig:MonelK500-Geom}b. This region spans a width of 0.5 mm and includes 280 grains, while the remaining part of the solid is modelled as an isotropic continuum without interfaces. In this way, the model can capture the two cracking modes observed in the experiments; in the absence of hydrogen, cracking occurred at the centre of the sample due to plastic instabilities, while in the presence of hydrogen, cracking took place near the notch tip and was of intergranular nature \cite{CS2020}. 

As in the experiments, the remote vertical displacement is applied at the top edge with a rate of $\dot{u}_y=1\times10^{-6} \ \text{mm}/\text{s}$. On the other hand, both vertical and horizontal displacements are constrained at the bottom edge. Also mimicking the experimental campaign, four Monel K-500 material heats were considered (Allvac, NRL LS, TR2 and NRL HS), with their macroscopic properties being given in Table \ref{tab:Heats-MonelK500}. The reader is referred to Ref. \cite{CS2020} for a comprehensive description on the different ageing and heat treatments employed for each lot. The phase field length scale is assumed to be equal to $\ell=0.025\ \text{mm}$ and the material diffusivity equals $D=1.3\times10^{-8}\ \text{mm}/\text{s}$ \cite{CS2020}. The remaining fracture properties for the bulk and interface are calibrated as follows. First, $G_c$ is chosen so as to reproduce the experimental force versus time response in air. Then, the cohesive interface parameters and the hydrogen damage coefficient $\chi$ are calibrated with the experiments conducted at the most aggressive conditions (applied potential of $E_A=-1.1$ V$_{\text{SCE}}$), and subsequently used to evaluate their predictive capabilities in other scenarios (different $E_A$ values). 

\begin{table}[H]
\centering
\caption{Material properties for the different heats of Monel K-500 considered.}
\begin{tabular}{@{}ccccc@{}}
\toprule
Heat & {$E \ (\text{GPa})$} & {$\nu$} & {$\sigma_{y}\ (\text{MPa})$} & n \\ \midrule
Allvac        & 180                       & 0.3            & 794.3                           & 0.058                       \\
NRL LS        & 198                       & 0.3            & 715.7                           & 0.054                       \\
TR2           & 202                       & 0.3            & 795                             & 0.055                       \\
NRL HS        & 191                       & 0.3            & 910.1                           & 0.050                       \\ \bottomrule
\end{tabular}
\label{tab:Heats-MonelK500}
\end{table}

Each Monel K-500 heat was tested in four different environments: in air (i.e., in the absence of a hydrogen-containing solution) and while being exposed to solutions with the applied potentials $E_A=-0.85$ V$_{\text{SCE}}$, $E_A=-0.95$ V$_{\text{SCE}}$, and $E_A=-1.1$ V$_{\text{SCE}}$. The diffusible hydrogen concentration associated with each charging condition was experimentally measured and used as input in the model - a prescribed $C_{\text{env}}$ magnitude at the surface. The values measured are given in Table \ref{tab:Potentials-MonelK500} (in ppm wt). 

\begin{table}[H]
\centering
\caption{Diffusible $C$ (wppm) for each Monel K-500 heat for each applied potential $E_A$.}
\begin{tabular}{@{}cccc@{}}
\toprule
Heat & {$E_{A}=-0.85\ \text{V}_{\text{SCE}}$} & {$E_{A}=-0.95\ \text{V}_{\text{SCE}}$} & {$E_{A}=-1.1\ \text{V}_{\text{SCE}}$} \\ \midrule
Allvac        & 1.9                             & 4.1                             & 7.5                            \\
NRL LS        & 1.3                             & 4.7                             & 18.3                           \\
TR2           & 3.7                             & 18.6                            & 26.2                           \\
NRL HS        & 4.7                             & 11.9                            & 23.4                           \\ \bottomrule
\end{tabular}
\label{tab:Potentials-MonelK500}
\end{table}

\begin{figure}[t]
	\centering
	\includegraphics[width=0.8\linewidth]{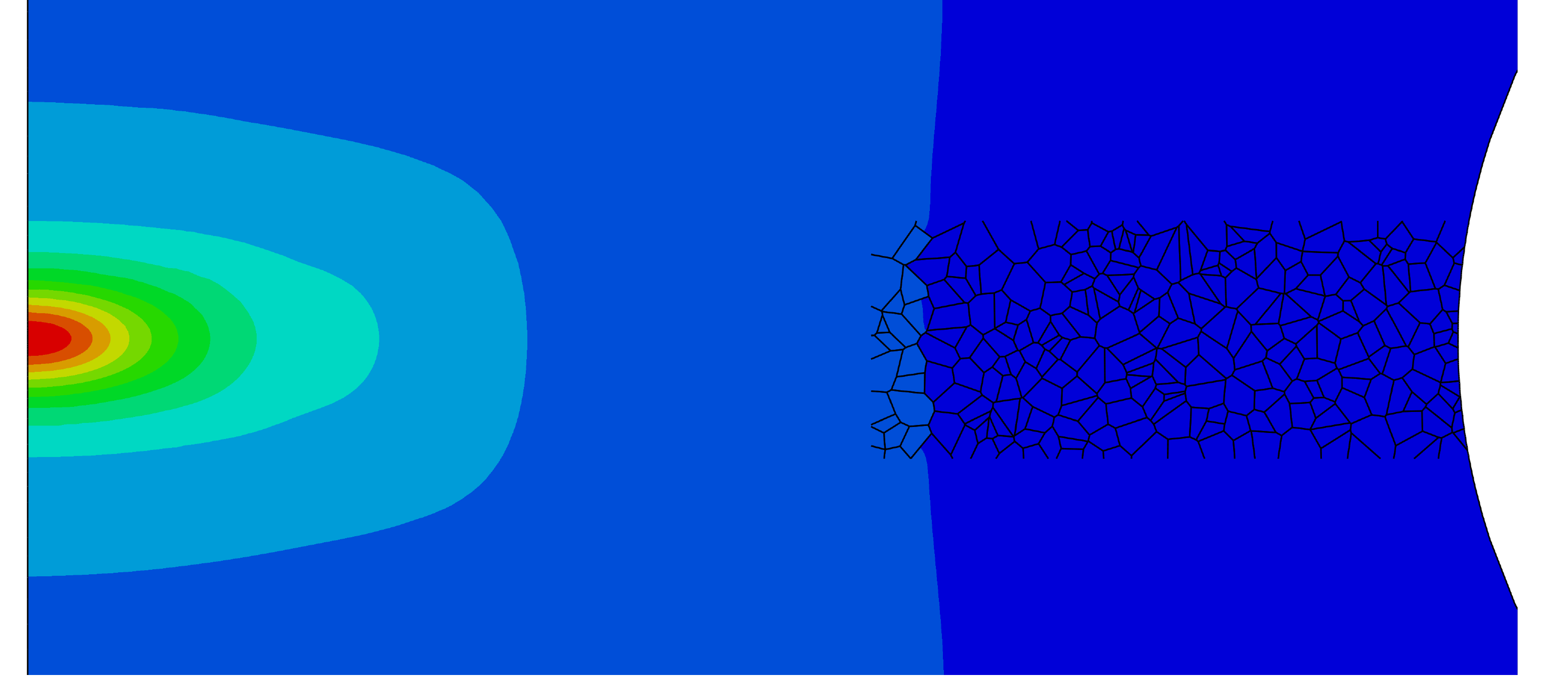} \\
	\includegraphics[width=0.2\linewidth]{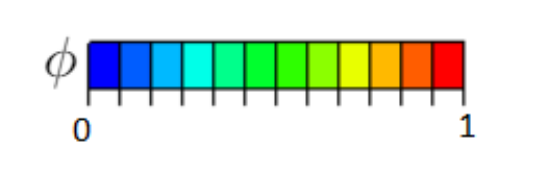}
	\caption{Virtual SSRT experiments on Monel K-500: phase field fracture contour at the time of crack initiation. In the absence of hydrogen, cracking takes place at the centre of the sample, in agreement with experimental observations.}
	\label{fig:TG-MonelK500}
\end{figure}

\begin{figure}[H]
\centering
\subfloat[][\label{fig:3.3.3a}]
{\includegraphics[width=0.5\linewidth]{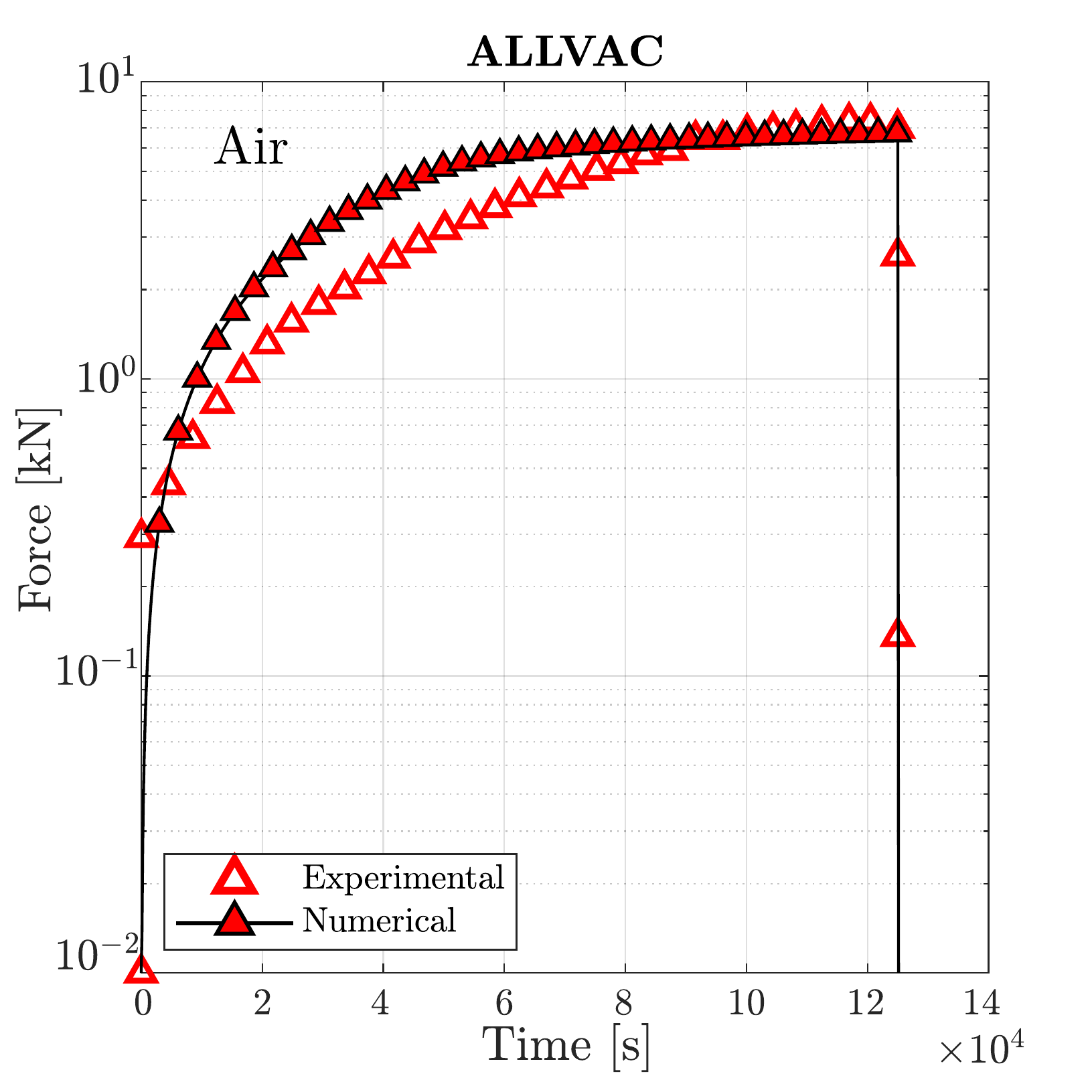}}
\subfloat[][\label{fig:3.3.3b}]
{\includegraphics[width=0.5\linewidth]{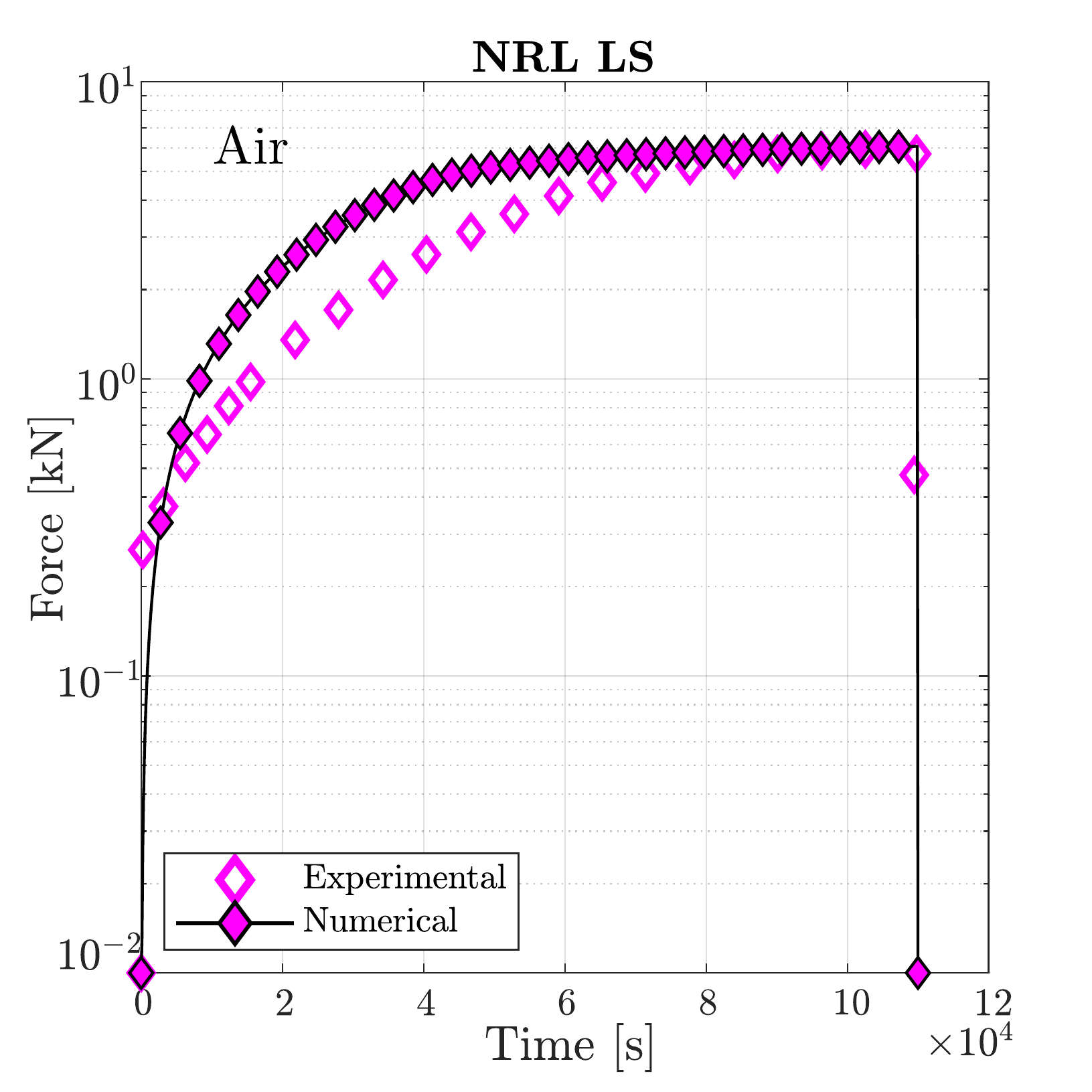}}
\\
\subfloat[][\label{fig:3.3.3c}]
{\includegraphics[width=0.5\linewidth]{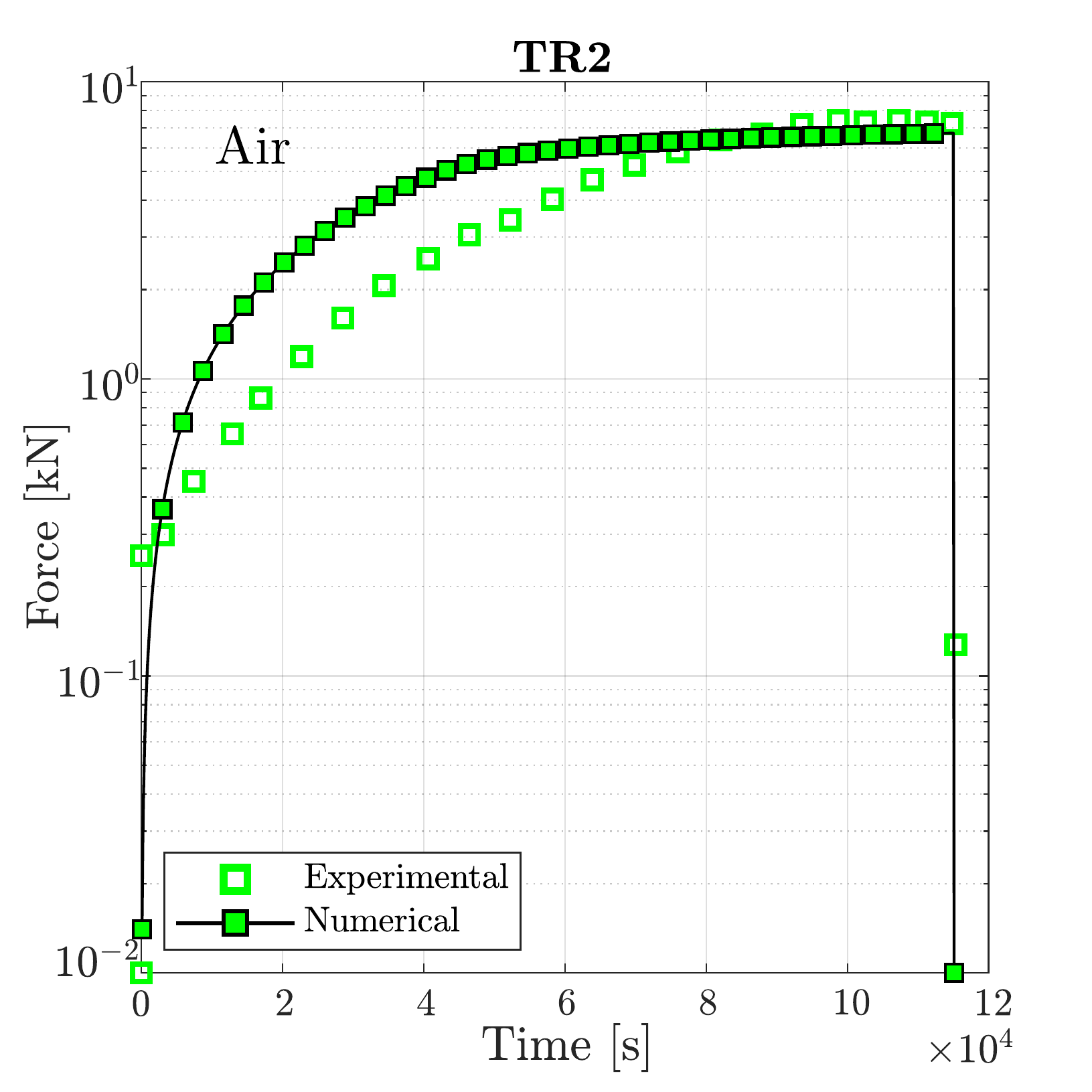}}
\subfloat[][\label{fig:3.3.3d}]
{\includegraphics[width=0.5\linewidth]{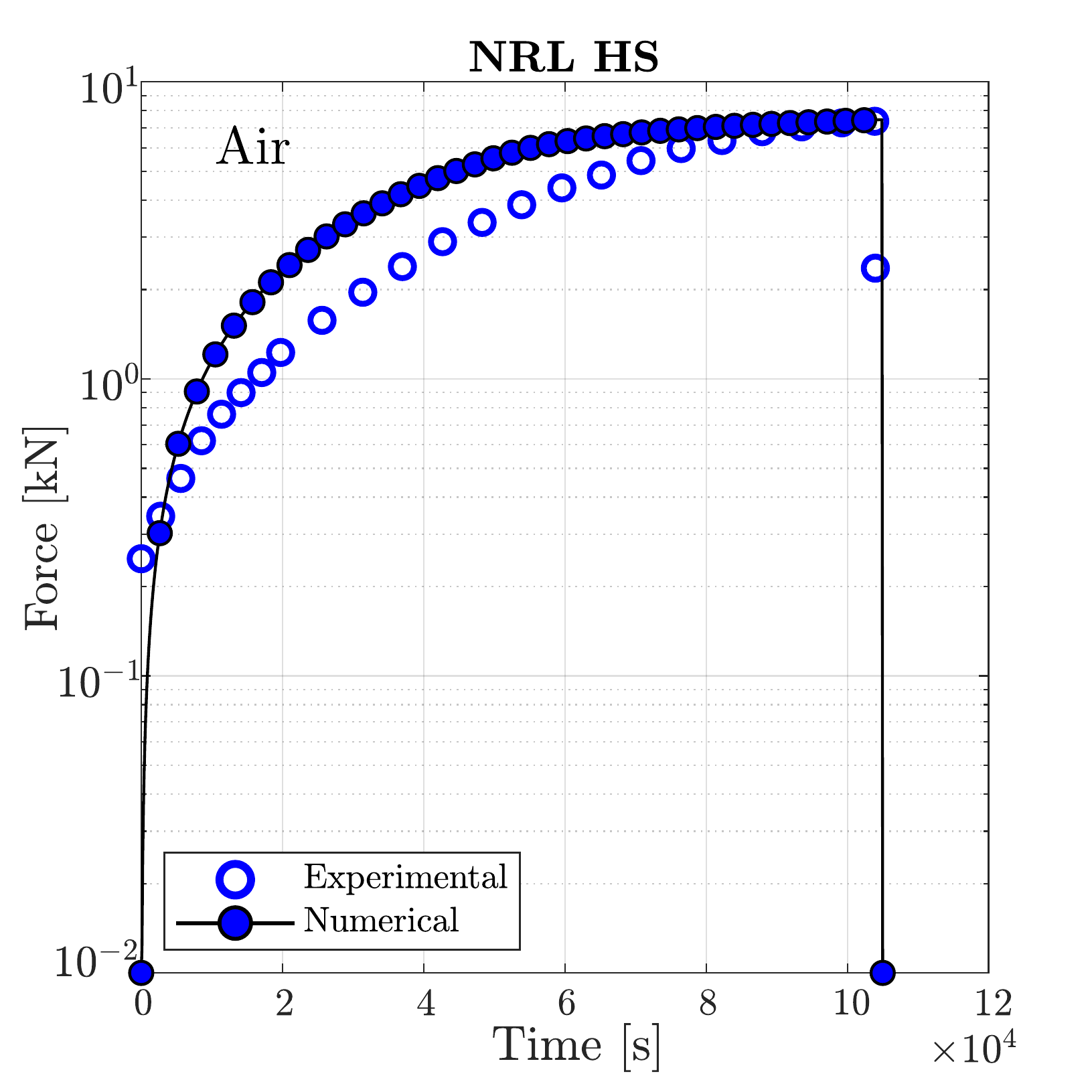}}
\\
\caption{Virtual SSRT experiments on Monel K-500: numerical and experimental force versus time curves obtained in the absence of hydrogen for each material lot; namely, (a) Allvac, (b) NRL LS, (c) TR2 and (d) NRL HS.}
\label{fig:Air-MonelK500-RFU}
\end{figure}

Simulations are first conducted in the absence of hydrogen. As shown in Fig. \ref{fig:TG-MonelK500}, damage initiates at the centre of the sample for sufficiently large loads, in agreement with experimental observations. The material toughness $G_c$ for each Monel K-500 lot is chosen so as to quantitatively reproduce the macroscopic force versus time curve, with the fitted values being 18.5, 18.1, 15.4 and 16.9 kJ/m$^2$ for, respectively, Allvac, NRL LS, TR2, and NRL HS. The force versus time curves obtained for each material lot are shown in Fig. \ref{fig:Air-MonelK500-RFU}. A good agreement is attained with experimental observations, with only small differences being observed in the early stages due to the additional compliance brought in by the machine displacement (an extensometer was not used \cite{CS2020}).\\ 

\begin{figure}[H]
	\centering
	\includegraphics[width=0.8\linewidth]{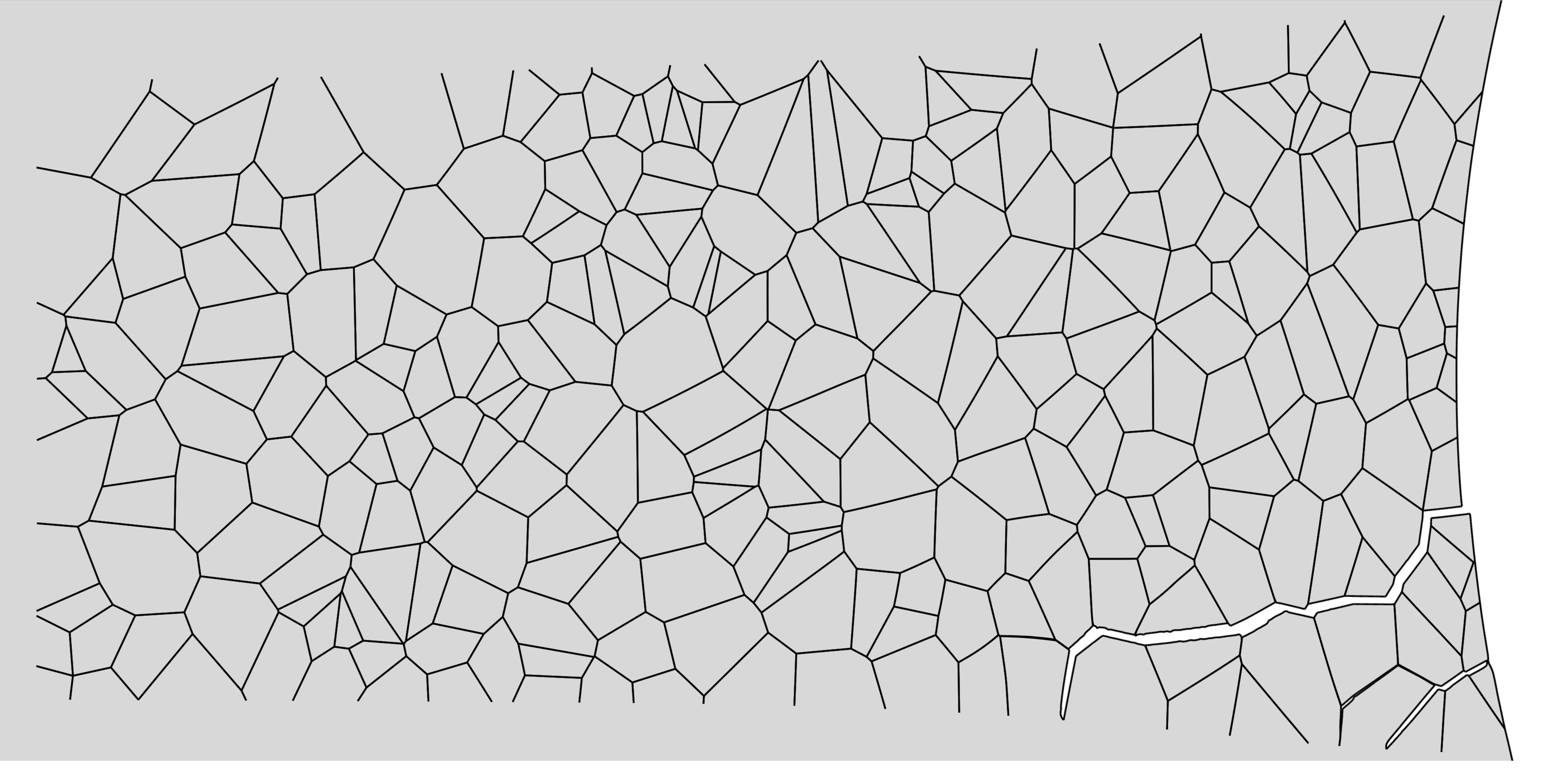}
	\caption{Virtual SSRT experiments on Monel K-500: representative result of intergranular cracking for a sample exposed to an applied potential of $E_A=-1.1$ V$_{\text{SCE}}$. A crack nucleates at a grain boundary adjacent to the notch tip and then propagates in-between grains towards the centre of the sample.}
	\label{fig:IG-MonelK500}
\end{figure}

\begin{figure}[H]
\centering
\subfloat[][\label{fig:3.3.5a}]
{\includegraphics[width=0.495\linewidth]{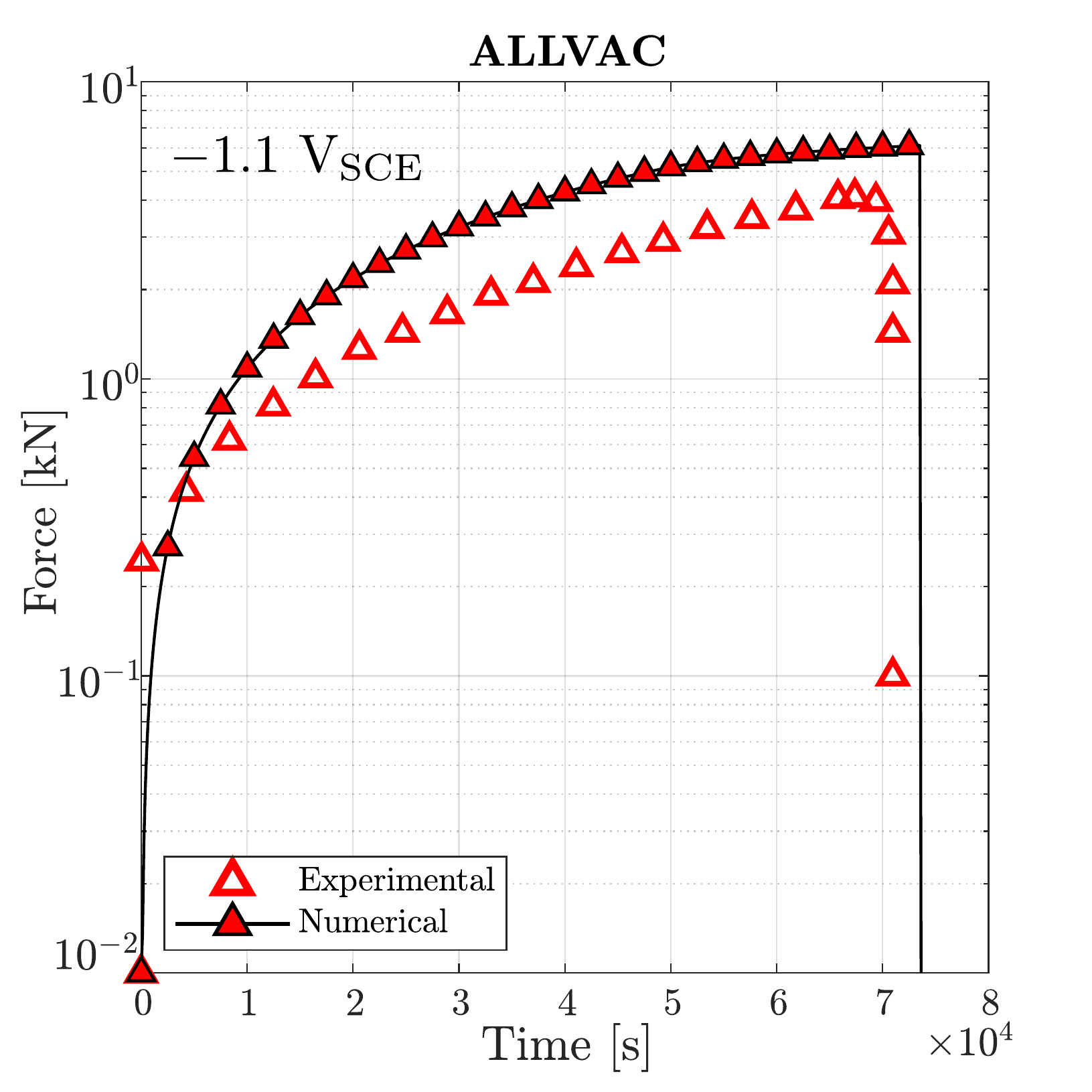}}
\subfloat[][\label{fig:3.3.5b}]
{\includegraphics[width=0.495\linewidth]{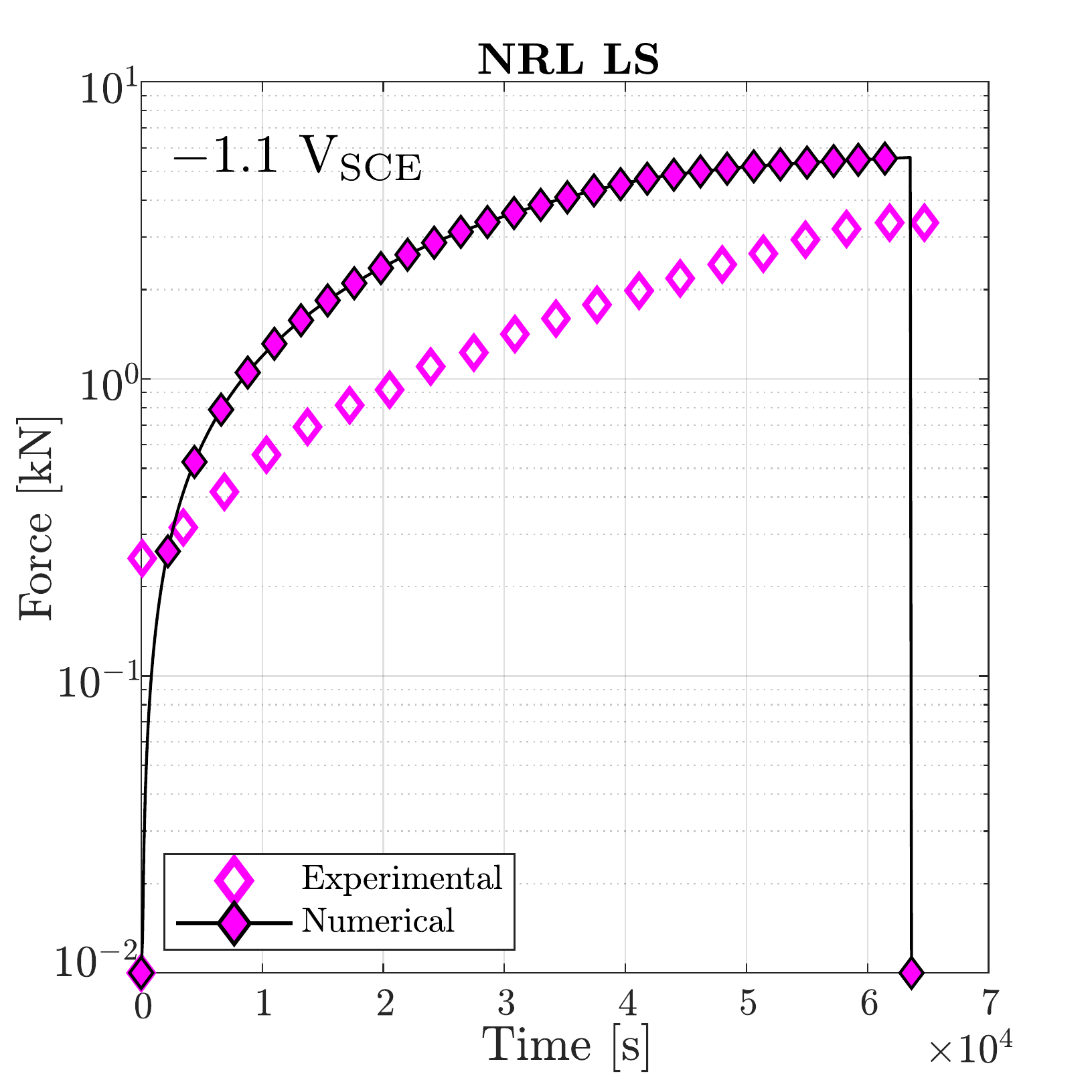}}
\\
\subfloat[][\label{fig:3.3.5c}]
{\includegraphics[width=0.495\linewidth]{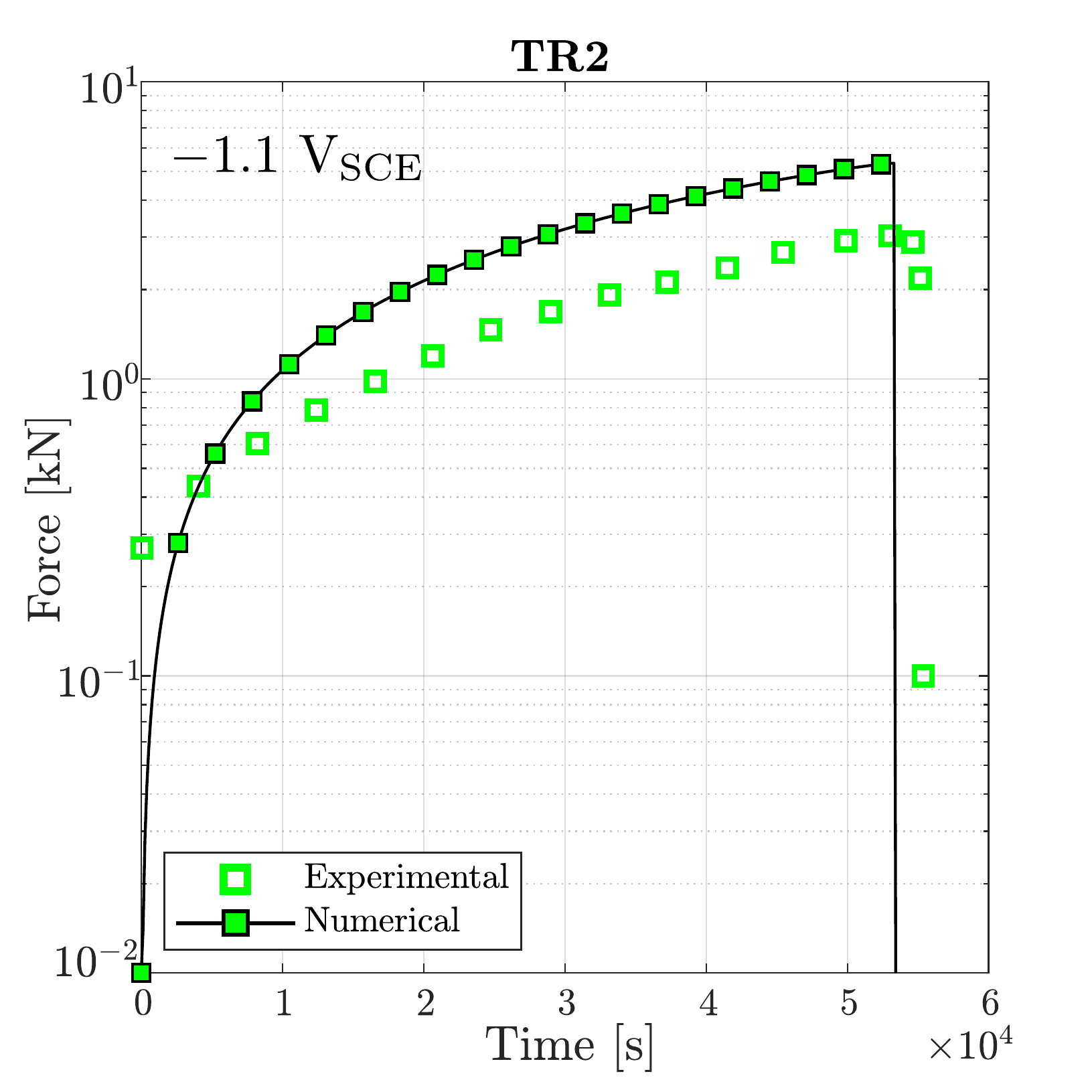}}
\subfloat[][\label{fig:3.3.5d}]
{\includegraphics[width=0.495\linewidth]{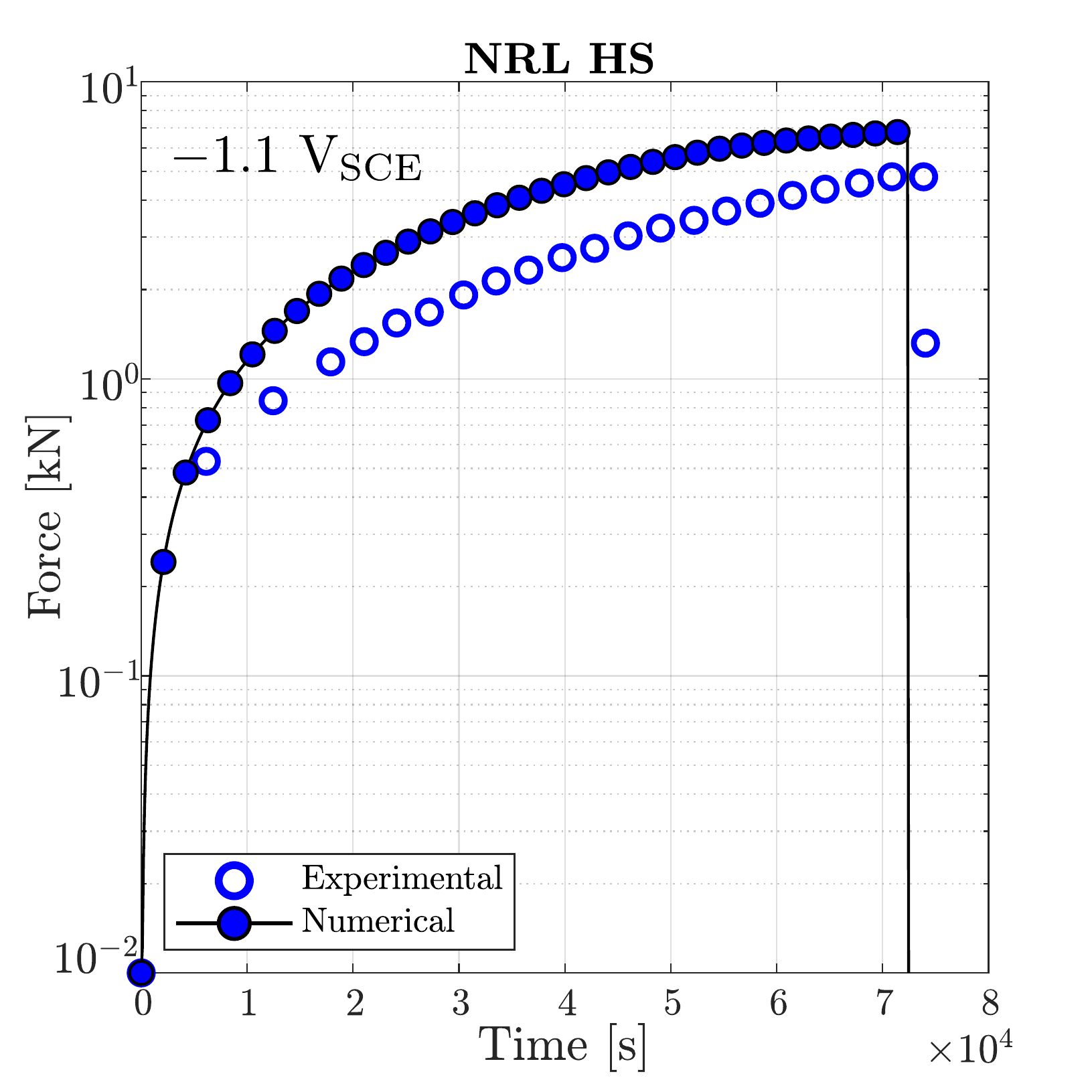}}
\\
\caption{Virtual SSRT experiments on Monel K-500: numerical and experimental force versus time curves obtained under an applied potential of $E_A=-1.1$ V$_{\text{SCE}}$ for each material lot; namely, (a) Allvac, (b) NRL LS, (c) TR2 and (d) NRL HS.}
\label{fig:1.1V-MonelK500-RFU}
\end{figure}

The samples with a higher degree of hydrogen uptake (those charged at $E_A=-1.1$ V$_{\text{SCE}}$) exhibit an intergranular fracture pattern, as shown in Fig. \ref{fig:IG-MonelK500}. A crack nucleates in a grain boundary adjacent to the notch tip and subsequently propagates along neighboring grain boundaries. The force versus time curves obtained for the four material lots under an applied potential of $E_A=-1.1$ V$_{\text{SCE}}$ are shown in Fig. \ref{fig:1.1V-MonelK500-RFU}. In all cases the failure is intergranular, starting at the notch tip and triggering a significant drop in the load carrying capacity before any noticeable increase in the phase field variable is observed. Thus, the calibrated model is also able to capture the ductile-to-brittle transition observed in this case study. 

Interestingly, the cracking event appears to occur rather suddenly, with the intergranular crack propagating into regions with low hydrogen content. This would suggest that the crack resulting from the decohesion of the grain boundaries exposed to a high hydrogen content would be sufficiently large to propagate in an unstable fashion through grain boundaries that have only been negligibly weakened by hydrogen. If that were to be the case, then these numerical results would suggest that SSRT is not compromised by early cracking and thus remains a valid test for measuring hydrogen embrittlement susceptibility. Additional, albeit limited, insight can be gained by a simple estimate of the transition flaw size $a_t=K_{Ic}^2/(\pi \sigma_Y^2)$. For Monel K-500 exposed to a relatively uniform hydrogen distribution resulting from an applied potential of $E_A=-1.1$ V$_{\text{SCE}}$, the transition flaw size would be on the order of 0.04 mm (see Table \ref{tab:Heats-MonelK500} and the $K_{TH}$ measurements of Ref. \cite{AM2016}). However, this quantity can increase to up to 4 mm in the absence of hydrogen. Thus, the magnitude of $a_t$ relevant to this scenario (a non-uniform distribution of hydrogen) falls between those two limiting cases, and could therefore be higher or lower than the crack extensions predicted (0.1-0.5 mm). Another source of uncertainty is the specific traction-separation law adopted, as assuming the existence of a damage dissipation region could add an additional source of fracture resistance.

The interface parameters that provide a quantitative agreement with experiments are given in Table \ref{tab:IntParam-MonelK500}. The damage coefficient $\chi$ was estimated based on previous (microstructurally-insensitive) phase field simulations \cite{CS2020}. The values of $\chi$ used are higher than those estimated using atomistic calculations for most common types of Ni grain boundaries \cite{Alvaro2015}. However, the choices of $\chi$ values are notably sensitive to the magnitude of the grain boundary binding energy considered in (\ref{eq:2.4.4}), and the estimation of this magnitude carries a degree of uncertainty \cite{Ai2013}. 

The calibrated model is used to reproduce the entire experimental campaign, conducting virtual experiments on the four material lots over the four environments considered. We summarise the outcome of the simulations in Table \ref{tab:FraMec-MonelK500}, indicating the failure model predicted (brittle integranular/IG or ductile transgranular/TG). Green check marks are used to denote when the predicted mode of cracking agrees with the experimental observation, with red crosses denoting otherwise. As observed, the model is capable of predicting the occurrence of hydrogen-assisted brittle failures in all but one case - the experiment on the NRL LS heat at $E_A=-0.85$ V$_{\text{SCE}}$. As discussed in Ref. \cite{CS2020}, this is a rare case as SEM images of the fracture region reveal intergranular features but the time to failure happens to be larger than that measured in the absence of hydrogen. It is thus concluded that the micromechanical model presented is capable of predicting hydrogen embrittlement upon appropriate calibration. 

\begin{table}[t]
\centering
\caption{Interface material parameters, as estimated by quantitatively reproducing the experiments conducted at an applied potential of $E_A=-1.1$ V$_{\text{SCE}}$.}
\begin{tabular}{@{}ccccc@{}}
\toprule
                                               & \textbf{Allvac}  & \textbf{NRL LS}  & \textbf{TR2}     & \textbf{NRL HS}  \\ \midrule
$k_n,k_t$ ($10^8 \ \text{MPa}/\text{mm}$)                          & $2.00$    & $2.00$    & $2.00$    & $2.00$    \\
$t_{nc,0},t_{tc,0}$ ($10^4 \ \text{MPa}$)                          & $2.07$ & $2.78$ & $2.23$ & $2.13$ \\
$\gamma_{IC,0},\gamma_{IIC,0}$ ($\text{kJ}/\text{m}^2$) & $1.07$           & $1.92$           & $1.24$           & $1.12$ \\
$\chi$ & 0.85             & 0.82             & 0.86          & 0.79          \\ \bottomrule
\end{tabular}
\label{tab:IntParam-MonelK500}
\end{table}

\begin{table}[H]
\centering
\caption{Virtual SSRT experiments on Monel K-500: predicted fracture mechanism (IG = intergranular, TG = transgranular) for each combination of material heat and environment. Green check marks and red crosses are used to respectively denote when predicted mode of cracking agrees or disagrees with the experimental observation.}
\begin{tabular}{@{}ccccc@{}}
\toprule
{\textbf{$\mathbf{E_A} \ (\text{V}_{\text{SCE}})$}} & \textbf{Allvac} & \textbf{NRL LS} & \textbf{TR2} & \textbf{NRL HS} \\ \midrule
0 (Air)              & TG \greencheck             & TG  \greencheck            & TG  \greencheck         & TG   \greencheck           \\
0.85                 & TG \greencheck             & TG \redmark              & TG \greencheck          & -               \\
0.95                 & TG \greencheck             & IG \greencheck             & IG \greencheck          & TG  \greencheck            \\
1.1                  & IG \greencheck             & IG \greencheck             & IG \greencheck          & IG \greencheck             \\ \bottomrule
\end{tabular}
\label{tab:FraMec-MonelK500}
\end{table}

\subsection{Failure of pure Ni samples at ambient and cryogenic temperatures}
\label{SubSec:Temp}

Finally, we employ the micromechanical cohesive zone - phase field formulation developed to shed light on the interplay between diffusion, deformation, temperature, and embrittlement on pure Ni. Harris \textit{et al.} \cite{Harris2018} investigated the contribution of mobile hydrogen-deformation interactions to hydrogen-induced intergranular cracking in polycrystalline Ni by testing hydrogen charged samples at both ambient (298 K) and cryogenic (77 K) temperatures. Their uniaxial mechanical tests showed that embrittlement (hydrogen-assisted intergranular cracking) occurred even at cryogenic temperatures, where dislocation-hydrogen interactions are precluded. This suggests that hydrogen-assisted decohesion of grain boundaries is a first-order mechanism in hydrogen embrittlement. It was also found that intergranular microcrack evolution was enhanced at room temperature, relative to 77 K, but a mechanistic interpretation of this finding was deemed complicated due to the multiple factors at play. Here, we examine the ability of our micromechanical model to quantitatively reproduce the seminal experiments by Harris \textit{et al.} \cite{Harris2018} and use the numerical insight provided to gain further understanding on the role of temperature in enhancing embrittlement. The material properties adopted correspond to those reported by Harris \textit{et al.} \cite{Harris2018}, which are listed in Table \ref{tab:MatProp-Nickel} as a function of the temperature and the environment. Two environmental conditions are considered: (i) samples tested in air, without any hydrogen pre-charging, and (ii) samples exposed to a hydrogen content of 4000 appm (79.5 wppm). In the latter, gaseous hydrogen charging is used and the hydrogen is distributed uniformly within the samples. Mimicking the experimental conditions, the uniaxial load is prescribed by applying a remote vertical displacement with a rate of $\dot{u}_y=0.0078$ mm/s, while the bottom edge is completely constrained ($u_x=u_y=0$).

\begin{table}[t]
\centering
\caption{Material properties reported for polycrystalline Ni samples at the two temperatures (77 and 298 K) and environments (79.5 wppm H and air) considered.}
\begin{tabular}{@{}ccccc@{}}
\toprule
\textbf{Temperature}                                                            & \multicolumn{2}{c}{\textbf{77 K}} & \multicolumn{2}{c}{\textbf{RT}} \\ \midrule
                                                & \textbf{no H}    & \textbf{H}    & \textbf{no H}    & \textbf{H}   \\
$E \ (\text{GPa)}$                                                                       & 227              & 227           & 202              & 202          \\
$\nu$                                                                           & 0.3              & 0.3           & 0.3              & 0.3          \\
$\sigma_{y} \  (\text{MPa})$                                                                & 222              & 233           & 182              & 192          \\
$n$                                                            & 0.159            & 0.159         & 0.140            & 0.140        \\
$D \ (\text{mm}^2/\text{s})$ & $10^{-15}$       & $10^{-15}$    & $10^{-9}$        & $10^{-9}$    \\ \bottomrule
\end{tabular}
\label{tab:MatProp-Nickel}
\end{table}

\begin{figure}[H]
\centering
\subfloat[][\label{fig:3.4.1a}]
{\includegraphics[width=0.25\linewidth]{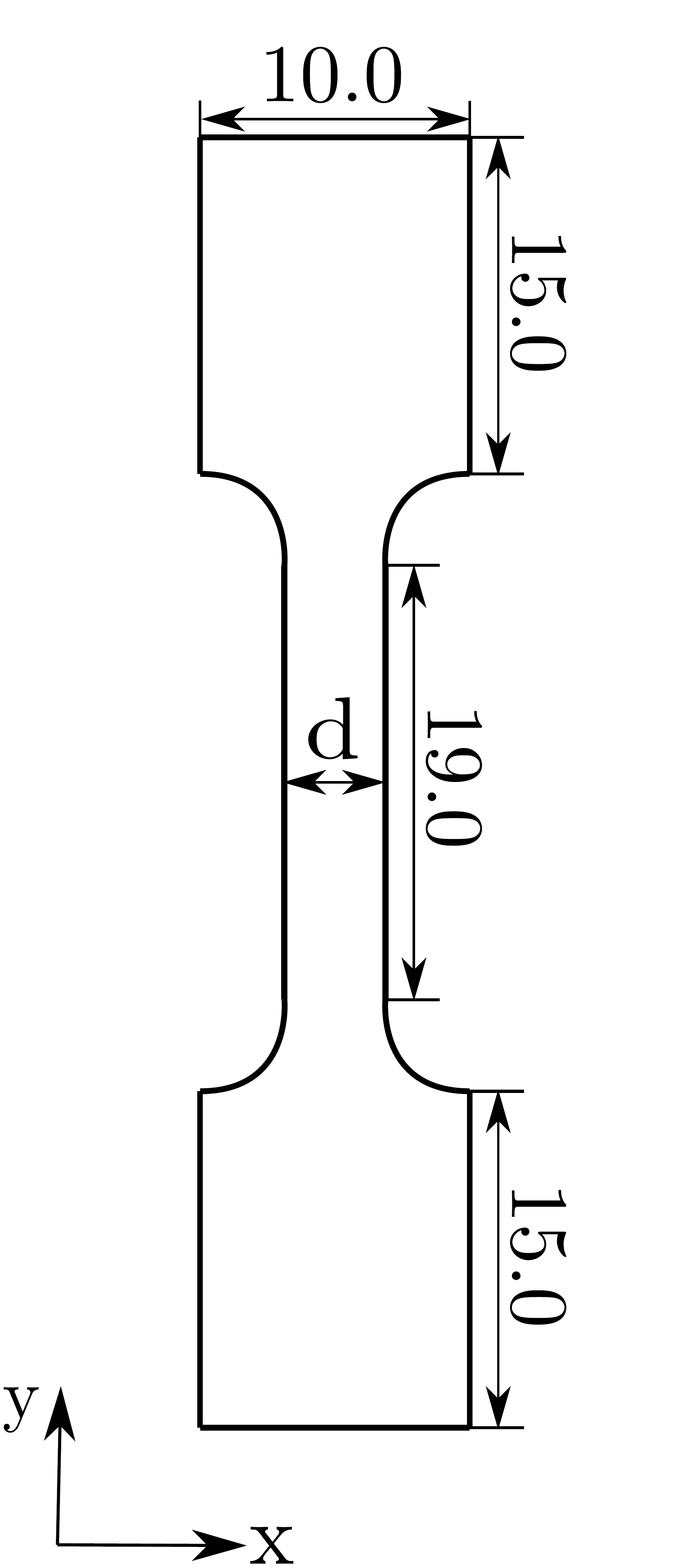}}
\hspace{0.1cm}
\subfloat[][\label{fig:3.4.1b}]
{\includegraphics[width=0.3\linewidth]{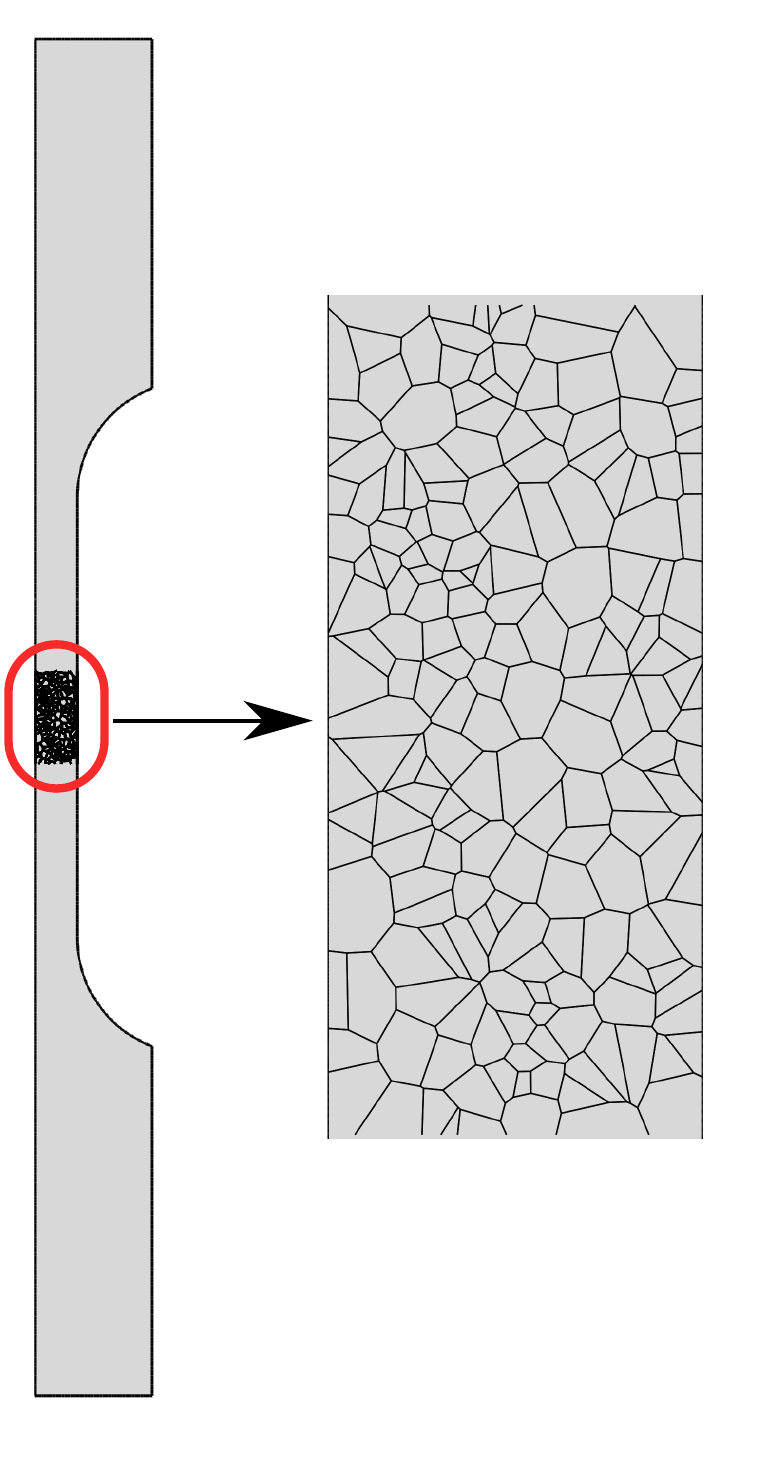}} 
\hspace{0.1cm}
\subfloat[][\label{fig:3.4.1c}]
{\includegraphics[width=0.25\linewidth]{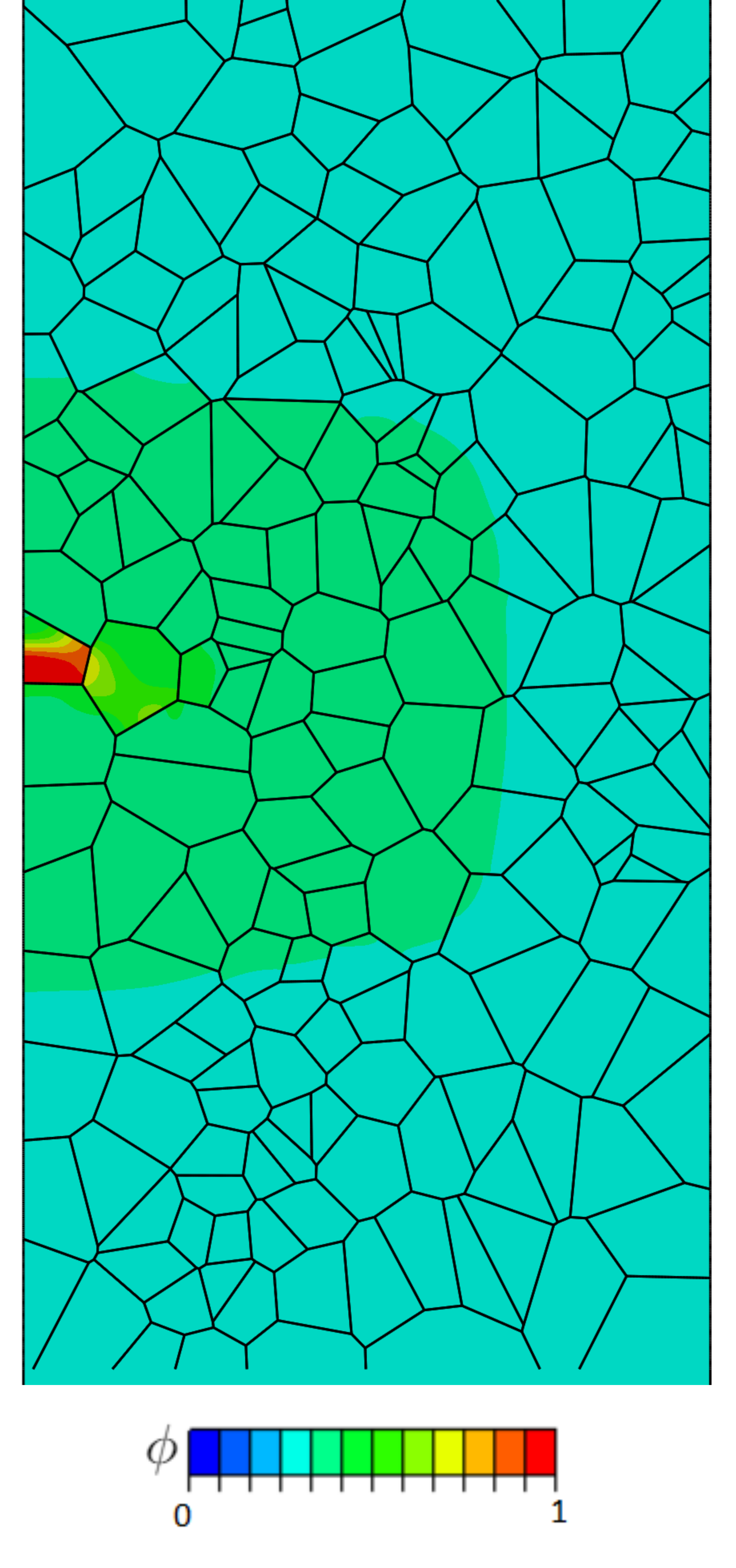}} \\
\subfloat[][\label{fig:3.4.1d}]
{\includegraphics[width=1\linewidth]{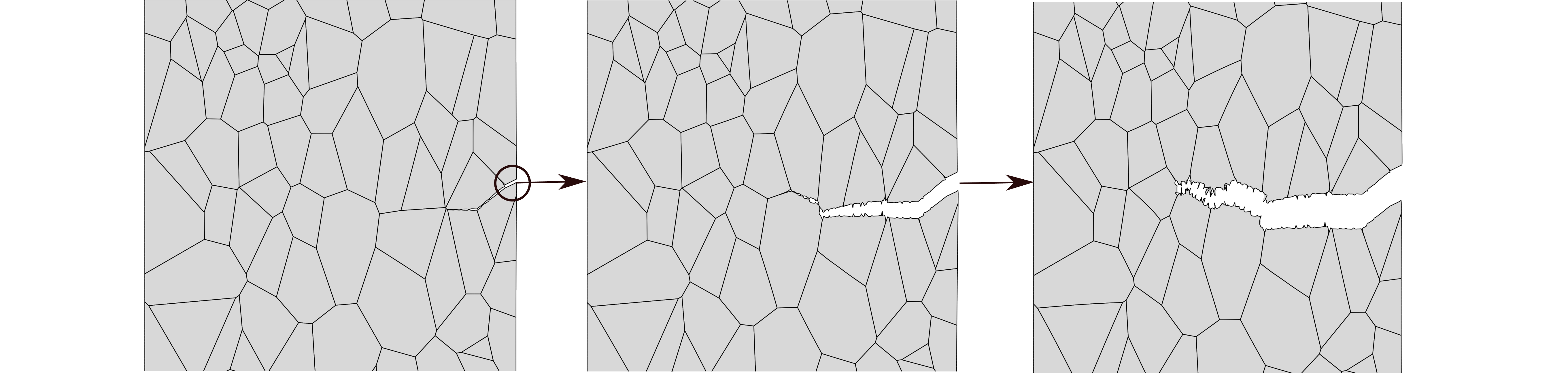}}
\caption{Role of temperature on the failure of polycrystalline Ni samples: (a) geometry with dimensions in mm, where d = 4 mm for 77 K and d = 3.6 mm for RT specimens \cite{Harris2018}, (b) augmented view of the microstructure region, with 200 grains, (c) ductile (transgranular) damage, as predicted by the phase field, and (d) intergranular crack nucleation and growth, as assisted by hydrogen. Representative results shown from calculations at ambient temperature.}
\label{fig:Geom-TG-IG-Nickel}
\end{figure}

The samples are cylindrical bars with the dimensions given in Fig. \ref{fig:Geom-TG-IG-Nickel}a. As in the previous case study, we take advantage of axial symmetry and model half of the 2D section using axisymmetric elements. A total of approximately 80,000 quadratic axisymmetric elements are used to discretise the model. As shown in Fig. \ref{fig:Geom-TG-IG-Nickel}b, we introduce a microstructural domain of 200 grains in the central region of the sample. The phase field length scale is taken to be equal to $\ell$ = 0.025 mm, and the toughness $G_c$ is calibrated to reproduce the experiments in the absence of hydrogen, rendering values of 4 kJ/m$^2$ (77 K) and 2.5 kJ/m$^2$ (RT). The interfacial cohesive properties are adjusted to reproduce the experiments at 77 K and then used to see if the results at room temperature can be predicted. The specific values used are given in Table \ref{tab:IntParam-Nickel} and, following the approach of Ref. \cite{IJP2021}, a phenomenological degradation law is adopted, such that
\begin{equation}
    \label{eq.3.4.1}
    \gamma_c (\theta_H) = \gamma_{C,0}\Big(17.52 \exp(-2.75\theta_H)\Big)
\end{equation}

\noindent with the Gibbs free energy being equal to 17 kJ/mol \cite{Choo1982}.

\begin{table}[t]
\centering
\caption{Calibrated traction-separation law parameters to describe the decohesion of pure Ni grain boundaries.}
\begin{tabular}{ccc}
\toprule
$\mathbf{k_n,k_t \ ( \text{\textbf{MPa}}/\text{\textbf{mm}})}$ & $\mathbf{t_{nc,0},t_{tc,0} \ (\text{\textbf{MPa}})}$ & $\boldsymbol{\gamma_{IC,0},\gamma_{IIC,0}} $  ($\text{\textbf{MPa}}\times\text{\textbf{mm}}$) \\ \midrule 
$2 \times 10^8$                                                           & $7.83 \times 10^4$                                         & $0.88$   \\ \bottomrule
\end{tabular}
\label{tab:IntParam-Nickel}
\end{table}

\begin{figure}[t]
	\centering
	\includegraphics[width=1\linewidth]{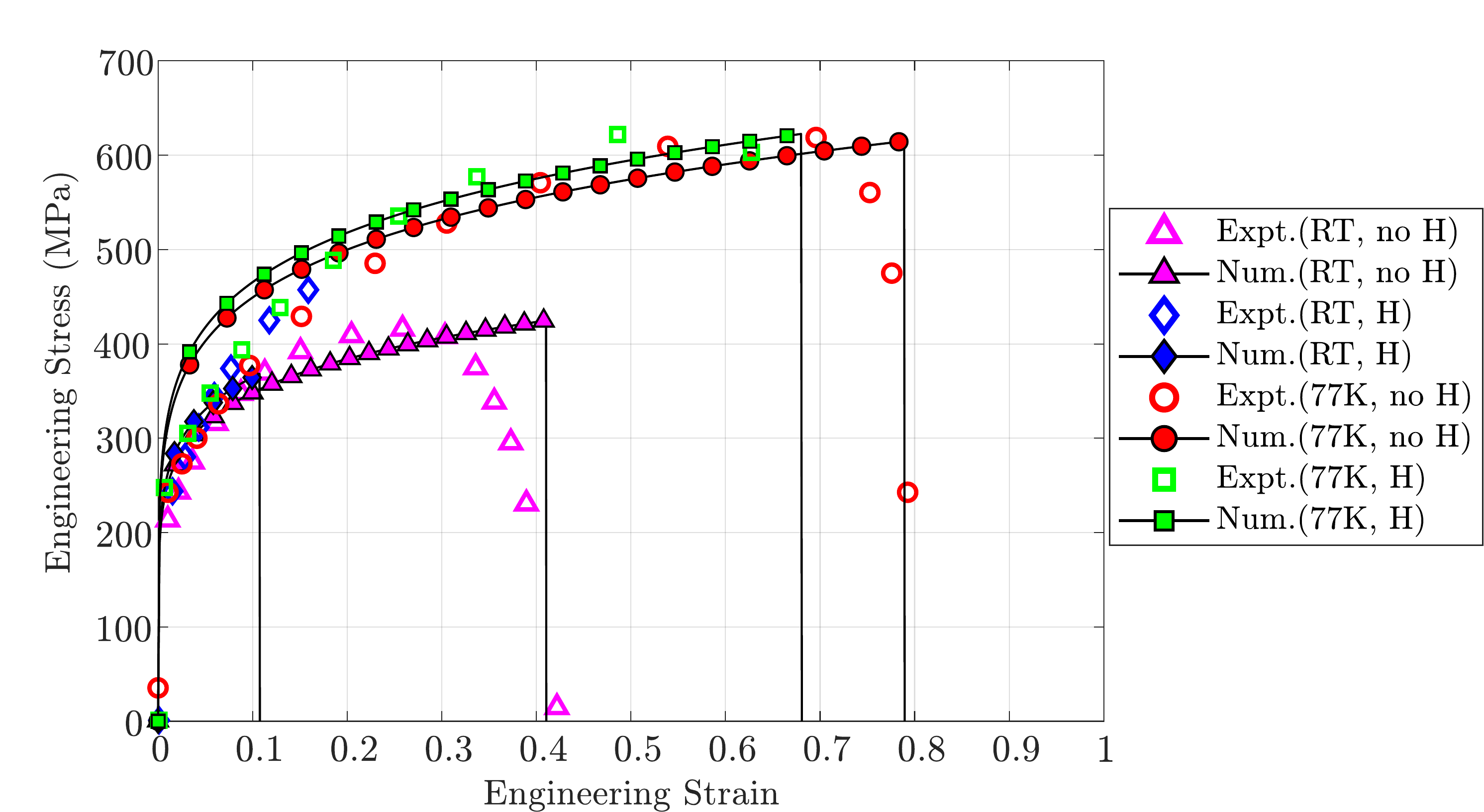}
	\caption{Role of temperature on the failure of polycrystalline Ni samples: engineering stress-strain curves for both H-charged and non-charged samples pulled to failure at 77K and at ambient temperature (298 K).}
	\label{fig:Nickel-RFU}
\end{figure}

As shown in Figs. \ref{fig:Geom-TG-IG-Nickel} and \ref{fig:Nickel-RFU}, the model is able to reproduce experimental measurements beyond the regimes of calibration, both qualitatively and quantitatively. Consider first the cracking patterns shown in Fig. \ref{fig:Geom-TG-IG-Nickel}. In the absence of hydrogen (Fig. \ref{fig:Geom-TG-IG-Nickel}c), failure takes place due to the onset of ductile (transgranular) damage in the centre of the sample, as predicted by the phase field order

\noindent parameter. However, when the sample is exposed to hydrogen, then cracking takes place in an intergranular fashion, as a result of the failure of the cohesive zone interfaces. The location of the grain boundary decohesion event that nucleates the brittle crack is random. For the microstructure and conditions of Fig. \ref{fig:Geom-TG-IG-Nickel}d, it occurs close to the edge of the sample, with the crack growing then both towards the outer surface and towards the centre of the sample. This change from ductile transgranular damage to brittle intergranular cracking due to hydrogen is observed at both 77 and 298 K, as in the experiments.

The quantitative results obtained for the four scenarios are shown in Fig. \ref{fig:Nickel-RFU}, in terms of the predicted and measured engineering stress-strain curves. A very good agreement with the experiments is observed. Interestingly, the good agreement observed for the case of the hydrogen-charged sample suggests that the higher degree of embrittlement observed at room temperature can be rationalised by the additional accumulation of hydrogen at grain boundaries due to diffusion, without the need for additional contributions from mechanisms such as those associated with hydrogen-deformation interactions.

\section{Summary and concluding remarks}
\label{Sec:ConcludingRemarks}

We have presented a new microstructurally-sensitive deformation-diffusion-fracture formulation to predict hydrogen embrittlement in elastic-plastic solids. The modelling framework is able to explicitly resolve the microstructure and capture the transition from ductile transgranular fracture to hydrogen-assisted brittle intergranular cracking. This is achieved by combining a phase field description of transgranular cracks with a cohesive zone model that simulates decohesion at the grain boundary interfaces. The capabilities of the model in bringing new insight and understanding are demonstrated by addressing three representative boundary value problems. First, the competition between transgranular and intergranular cracking is investigated with a single edge notched tension specimen that contains 50 grains. The expected qualitative trends are captured, with cracking mechanisms changing as a function of the environment. Second, recent slow strain rate test (SSRT) experiments on a Ni-Cu superalloy (Monel K-500) \cite{CS2020} are simulated to demonstrate the ability of the model to quantitatively predict failure times for different environments and material heats. The model is also used to discuss the suitability of SSRT experiments, showing that cracks nucleating along grain boundaries located near the surface can rapidly propagate inwards and significantly reduce the load carrying capacity. Finally, the paradigmatic experiments by Harris \textit{et al.} \cite{Harris2018} on polycristalline Ni samples at ambient and cryogenic temperatures are reproduced. The model is shown to predict both the qualitative cracking modes and the quantitative stress-strain responses for the four conditions (with and without hydrogen, at 77 K and 298 K). Furthermore, mechanistic insight into the embrittlement of polycrystalline Ni is gained, showing that grain boundary decohesion is a first order effect, and that the differences between the responses observed at ambient and cryogenic temperatures can be rationalised by the additional content of hydrogen accumulated in grain boundaries due to diffusion. The numerical experiments conducted also showcase the computational robustness of the method, with significant cracking being predicted without convergence issues.

\section*{Acknowledgements}
\label{Sec:Acknowledgeoffunding}

A. Valverde-Gonz\'alez acknowledges the financial support from Erasmus+ funding (Project 2020-1-IT02-KA103-078114) for his visiting time in University of Seville during the period 15/06-15/09 2021. E. Mart\'{\i}nez-Pa\~neda acknowledges financial support from the EPSRC [grant EP/V009680/1] and from UKRI's Future Leaders Fellowship programme [grant MR/V024124/1]. A. Quintanas-Corominas acknowledges financial support from the European Union-NextGenerationEU and the Ministry of Universities and Recovery, Transformation and Resilience Plan of the Spanish Government through a call of the University of Girona (grant REQ2021-A-30). J. Reinoso acknowledges financial support from the Ministry of Science, Innovation and Universities (Project PGC2018-099197-B-I00), the Consejer\'{\i}a de Econom\'{\i}a y Conocimiento, Junta de Andaluc\'{\i}a, and the European Regional Development Fund (Projects US-1265577 and P20-00595). M. Paggi acknowledges financial support from the Italian Ministry of Education, University and Research (MIUR) to the research project of relevant national interest (PRIN 2017) “XFAST-SIMS: Extra fast and accurate simulation of complex structural systems” (CUP: D68D19001260001).






\bibliographystyle{elsarticle-num}
\bibliography{library}

\end{document}